\documentclass[sn-basic, twocolumn]{sn-jnl}

\usepackage{graphicx}%
\usepackage{multirow}%
\usepackage{amsmath,amssymb,amsfonts}%
\usepackage{amsthm}%
\usepackage{mathrsfs}%
\usepackage[title]{appendix}%
\usepackage{xcolor}%
\usepackage{textcomp}%
\usepackage{manyfoot}%
\usepackage{booktabs}%
\usepackage{algorithm}%
\usepackage{algorithmicx}%
\usepackage{algpseudocode}%
\usepackage{listings}%

\usepackage{siunitx}
\newcommand{\Aspot}{A_\mathrm{spot}}
\newcommand{\AWf}{\omega_f}
\newcommand{\AWi}{\omega_\circ}
\newcommand{\AWr}{\omega_\mathrm{r}}
\newcommand{\Bspot}{B_\mathrm{spot}}
\newcommand{\Bsw}{B_\mathrm{SW}}
\newcommand{\CD}{C_\mathrm{d}}
\newcommand{\Dlat}{\Delta\mathrm{lat}}
\newcommand{\Dlon}{\Delta\mathrm{lon}}
\newcommand{\Drot}{\Delta\mathrm{rot}}
\newcommand{\for}{\mathrm{for} \hs}
\newcommand{\Fsct}{F_{\star_\mathrm{ctr}}}
\newcommand{\Fspot}{\Phi_\mathrm{spot}}
\newcommand{\hs}{\;\;\;\;}
\newcommand{\inci}{\mathrm{inc}_\circ}
\newcommand{\Ispot}{I_\mathrm{spot}}
\newcommand{\LAspt}{\mathrm{lat_{spot}}}
\newcommand{\lati}{\mathrm{lat}_\circ}
\newcommand{\loni}{\mathrm{lon}_\circ}
\newcommand{\LOspt}{\mathrm{lon_{spot}}}
\newcommand{\Mcme}{M_\mathrm{CME}}
\newcommand{\Npol}{N_\mathrm{pol}}
\newcommand{\Ntor}{N_\mathrm{tor}}
\newcommand{\Porb}{P_\mathrm{orb}}
\newcommand{\rAP}{r_\mathrm{a,p}}
\newcommand{\rGA}{r_\mathrm{g,a}}
\newcommand{\rM}{r_\mathrm{M}}
\newcommand{\Rp}{R_\mathrm{p}}
\newcommand{\Rspot}{R_\mathrm{spot}}
\newcommand{\Rss}{R_\mathrm{SS}}
\newcommand{\Teff}{T_\mathrm{eff}}
\newcommand{\vr}{v_\mathrm{r}}
\newcommand{\vlon}{v_\mathrm{lon}}
\let\mc\multicolumn

\chardef\us=`\_

\begin{document}

\title[Trajectories of CMEs from Solar-type Stars]{Trajectories of Coronal Mass Ejection from Solar-type Stars}


\author*[1,2]{\fnm{Fabian} \sur{Menezes}}\email{menezes.astroph@gmail.com}
\author[1]{\fnm{Adriana} \sur{Valio}}\email{avalio@craam.mackenzie.br}
\author[3]{\fnm{Yuri} \sur{Netto}}
\author[1]{\fnm{Alexandre} \sur{Araújo}}
\author[4]{\fnm{Christina} \sur{Kay}}\email{christina.d.kay@nasa.gov}
\author[5]{\fnm{Merav} \sur{Opher}}

\affil*[1]{\orgdiv{Centro de R{\'a}dio Astronomia e Astrof{\'i}sica Mackenzie (CRAAM)}, \orgname{Universidade Presbiteriana Mackenzie}, \orgaddress{
\city{S{\~a}o Paulo}, 
\country{Brazil}}}

\affil[2]{\orgdiv{Laboratoire Lagrange}, \orgname{Université Côte d'Azur, Observatoire de la Côte d'Azur, CNRS}, \orgaddress{
\city{Nice}, 
\country{France}}}

\affil[3]{\orgdiv{Instituto de Astronomia, Geofísica e Ciências Atmosféricas (IAG), Departamento de Astronomia}, \orgname{Universidade de São Paulo}, \orgaddress{
\city{S{\~a}o Paulo}, 
\country{Brazil}}}

\affil[3]{\orgdiv{Heliophysics Science Division}, \orgname{NASA Goddard Space Flight Center}, \orgaddress{
\city{Greenbelt}, 
\country{USA}}}

\affil[3]{\orgdiv{Department of Astronomy}, \orgname{Boston University}, \orgaddress{
\city{Boston}, 
\country{USA}}}

\abstract{The Sun and other solar-type stars have magnetic fields that permeate their interior and surface, extends through the interplanetary medium, and is the main driver of stellar activity. Stellar magnetic activity affects physical processes and conditions of the interplanetary medium and orbiting planets. Coronal mass ejections (CMEs) are the most impacting of these phenomena in near-Earth space weather, and consist of plasma clouds, with magnetic field, ejected from the solar corona. Precisely predicting the trajectory of CMEs is crucial in determining whether a CME will hit a planet and impact its magnetosphere and atmosphere. Despite the rapid developments in the search for stellar CMEs, their detection is still very incipient. In this work we aim to better understand the propagation of CMEs by analysing the influence of initial parameters on CME trajectories, such as position, velocities, and stellar magnetic field's configuration. We reconstruct magnetograms for Kepler-63 (KIC 11554435) and Kepler-411 (KIC 11551692) from spot transit mapping, and use a CME deflection model, ForeCAT, to simulate trajectories of hypothetical CMEs launched into the interplanetary medium from Kepler-63 and Kepler-411. We apply the same methodology to the Sun, for comparison. Our results show that in general, deflections and rotations of CMEs decrease with their radial velocity, and increase with ejection latitude. Moreover, magnetic fields stronger than the Sun's, such as Kepler-63's, tend to cause greater CME deflections.}

\keywords{Sun: coronal mass ejections (CMEs) -- planet-star interactions -- stars: activity -- stars: magnetic fields -- stars: solar-type -- starspots}



\maketitle



\section{Introduction} \label{sec:intro}

The magnetic field in solar-type stars is the main driver of phenomena related to stellar activity. Such phenomena occur from the photosphere (e.g. spots) throughout the corona (e.g. flares, eruptions, proeminences) and can extend to outer layers and over the interplanetary medium, as an ionised wind and coronal mass ejection (CME). These products of solar activity also influence physical processes and conditions in the Earth's atmosphere and in the near-Earth interplanetary medium, which is known as space weather.

The events that most impact space weather are CMEs -- clouds of plasma ejected from the solar corona, with a magnetic field stronger than the background interplanetary magnetic field of the solar wind \citep{gopalswamy09}. Despite Earth being 1~AU from the Sun, CMEs can be geo-effective, that is, can hit Earth and affect the geomagnetic field and its atmosphere, depending on the solar coordinates where they are ejected from. When they reach Earth, they can cause space weather phenomena, such as auroras, which pose no risk and are considered a beautiful natural spectacle. On the other hand, the collision of a CME with the Earth's magnetosphere can accelerate particles and alter the Earth's electric and magnetic fields \citep{baker98}, which can drive geomagnetic storms, damage satellites, communication systems, transmission networks and electrical distribution, and even endanger the lives of astronauts.

However, space weather in other stellar systems can differ significantly from that of the Solar System. Exoplanets can orbit very close to their star, and stellar activity and magnetic field strength can be vastly greater compared to that of the Sun. In these cases, orbiting exoplanets may be subject to more extreme space weather conditions compared to Earth. The action of intense UV flux, stellar wind, and the frequent impact of CMEs can cause the atmospheric erosion of a planet. This may result in a surface exposed to ionising radiation, which is harmful to biological organisms \citep{estrela18, estrela20}.

The CME trajectory is a very relevant factor in its potential impact on a planet. In the 1970s, with the first CME observations from the Sun, deflections in their trajectories were observed \citep{macqueen86}, and continued to be observed since then \citep{byrne10}. CMEs originating at high latitude tend to be deflected by coronal holes towards the equator, which can cause a CME that was initially expected to hit Earth, not to reach Earth and vice versa \citep{byrne10, mays15, mostl15}. Magnetic forces -- magnetic tension and magnetic pressure gradient -- have become a well-accepted explanation for the cause of CME deflections \citep{gopalswamy09, gui11, kay15b}. In general, CMEs tend to drift to local null points of magnetic field due to magnetic forces in the lower corona \citep{cremades04, kilpua09, gopalswamy09}. That is, they move away from the coronal holes and are directed towards the Heliospheric Current Sheet -- regions of minimal magnetic energy.

Furthermore, CME rotations have been reported both in observational studies \citep{vourlidas11, nieves12} and in studies with numerical simulations \citep{torok03, fan04, lynch10}. Rotation in a CME changes the orientation of its magnetic field. Knowing the direction of the magnetic field of CMEs increases the accuracy in predicting its impact \citep{gonzalez99}. A northbound field tends not to generate large changes in the geomagnetic field, since both have the same direction, which tends to generate only compression in the Earth's magnetic field. However, a southbound field, when colliding with the geomagnetic field, can generate great changes, as it initially generates compression and then reconnection between the lines of both fields, since they have opposite directions.

In the extrasolar scenario, there are studies on the probability of CMEs hitting exoplanets \citep{kay16, kay19}. For exoplanets orbiting M dwarfs, the probability is 0.5 to 5 CME impacts per day; for hot Jupiters orbiting solar-like stars, the probability is 0.05 to 0.5 CME impacts per day. By combining knowledge of Solar System space weather with properties inferred from other systems, it was possible to study the trajectory of hypothetical CMEs launched into the interplanetary medium.

\cite{kay16} reported that the probability of impact decreases with increasing inclination of the planetary orbit in relation to the Astrospheric Current Sheet (stellar analogue of the Heliospheric Current Sheet). \cite{kay19} simulated CMEs from the young Sun’s twin, $\kappa^1$ \textit{Ceti}, to study the effects of deflections in the magnetic corona of the young Sun and their impact on the early Venus, Earth, and Mars. Their work suggests that CMEs tend to propagate within a small cone about the ecliptic plane increasing the impact frequency of CMEs with planetary magnetospheres near this plane to $\sim$30~\%.

Precisely predicting the trajectory of CMEs is crucial in determining whether a CME will hit a planet and impact its magnetic field and atmosphere. Knowing the deflection and rotation of a solar CME allows the prediction of Earth impacts, which provides an opportunity to mitigate the negative effects produced by physical processes that affect the Earth and nearby space. The prediction with numerical models is crucial for a better understanding of phenomena, such as CMEs, both in solar space weather and in extrasolar cases. However, little is known about CMEs from other solar-type stars. Despite the rapidly development of searches for stellar CMEs, they are still not well constrained \citep[][and references therein]{argiroffi19, cliver22}. So far, probable detections of stellar CMEs have been presented, however their physical parameters, which are difficult to access from observations, have not been determined for the majority of known events \citep{leitzinger22}.

In this work we aim to better understand the propagation of CMEs by analysing the influence of some of their parameters and the stellar magnetic field on CME trajectories. We use a CME deflection model -- Forecasting a CME’s Altered Trajectory \citep[ForeCAT\footnote{ForeCAT is available at \url{https://github.com/ckay314/ForeCAT}};][]{kay13, kay15b} -- to simulate trajectories of hypothetical CMEs launched into the interplanetary medium from the Sun and also from the stars Kepler-63 and Kepler-411, with different input parameters. These solar-type stars have had their spotted surface mapped from planetary transits at different latitudes \citep{netto20, araujo21}.
Then, we analyse how initial position, velocities, and strength and configuration of magnetic field influence the trajectory of CMEs.

The next section (\ref{sec:stars}) describes the two stars, besides the Sun, from which we simulate CMEs, whereas Section~\ref{sec:model} describes the the magnetic background used in the model, as well as the input parameters of the CMEs and the stars. Section~\ref{sec:results} presents the results of the simulations, that is the variations of longitude, latitude, and rotation over the CME trajectories. Finally, Section~\ref{sec:conclusion} presents the conclusions.
\section{Kepler Stars} \label{sec:stars}

The stellar age plays a key role in the stellar magnetic field strength and configuration, rotation, activity, and mass loss \citep{skumanich72, wood02, charbonneau14, morris20}. However, despite their young age in this work we consider Kepler-63 and Kepler-411 to be similar enough to the Sun. The following parameters and features (described in detail in Section~\ref{sec:model}) are set to be the same as those of the Sun: density and velocity of the stellar wind, the CME's shape, size, expansion, mass and velocity, the source-surface height of the background magnetic field, and the relations between intensity and magnetic field of starspots. Thus, the free parameters that influence the trajectory of CMEs are: initial position, velocities, and strength and configuration of the magnetic field.

Kepler-63 and Kepler-411 are known to exhibit strong stellar magnetic activity \citep{sanchis13, estrela16, sun19, netto20, araujo21}. Kepler-63 is a young G-type star ($210~\pm~35$~Myr), with effective temperature, $\Teff~=~5576~\pm~50$~K, mass, $M_\star~=~0.984_{-0.04}^{+0.035}~M_\odot$, radius, $R_\star~=~0.901~_{-0.022}^{+0.027}~R_\odot$, and rotation period, $P_\star~=~5.401~\pm~0.014$~days \citep{sanchis13}. The orbiting gas giant planet, Kepler-63~b, has an orbital period, $\Porb~\simeq~9.43$~days \citep{sanchis13}, a radius, $\Rp~=~0.0644~R_\star$ ($\sim$6.3~$R_\oplus$), and an semi-major axis of 19.35~$R_\star$ \citep{netto20}. Moreover, the planet is in an almost polar orbit with respect to the rotation axis of the star \citep{sanchis13}, which makes it possible to map spots at several latitudes. Kepler-63 has magnetic activity cycle period of 1.27 years, \textit{i.e.} about 10 times shorter than the solar cycle \citep{estrela16}.

Kepler-411 is a young solar-type K2V star ($212~\pm~31$~Myr) which shows features that indicate relatively strong magnetic activity \citep{sun19}. Its mass and radius are $M_\star~=~0.83_{-0.10}^{+0.04}~M_\odot$ and $R_\star~=~0.79~_{-0.06}^{+0.07}~R_\odot$, respectively \citep{wang14}. This star hosts at least 4 planets \citep{sun19}, 3 of which transit the star with orbits projected at \ang{-11} (Kepler-411~b), \ang{-21} (Kepler-411~c) and \ang{-49} (Kepler-411~d) of the stellar latitude \citep{araujo21}. The orbital periods of Kepler-411~b, Kepler-411~c, Kepler-411~d are respectively $\Porb~=~3.0051~\pm~0.00005$~days, $\Porb~=~7.834435~\pm~0.000002$~days and $\Porb~=~58.02~\pm~0.0002$~days \citep{wang14}, and their radii are respectively $\Rp~=~0.024~\pm~0.002~R_\star$, $\Rp~=~0.042~\pm~0.002~R_\star$ and $\Rp~=~0.040~\pm~0.002~R_\star$ \citep{araujo21}. \cite{araujo21} measured the star's differential rotation and determined an average rotation period of $P_\star~=~10.52$~days.
\section{ForeCAT Model Setup} \label{sec:model}

\subsection{Magnetic Background} \label{sec:model_magnetic}

\cite{kay13, kay15b} developed the ForeCAT model that calculates the deflection and rotation of a CME due to tension and pressure gradient of the solar magnetic field. We use an older, slightly less complex version of ForeCAT than the one presented in \cite{kay22}, since we do not consider an elliptical CME cross section for this work, given how little observational constraints there are on extrasolar CMEs. The model calculates the three-dimensional (3D) trajectories of CMEs and reproduces the general trends of observed CMEs. Input parameters to the model are the characteristics of the CME and the star, as well as magnetograms.

The simulated CME is embedded in a magnetic background that determines its deflection and rotation, causing changes in the latitude, longitude, and orientation of the torus as it propagates outward \citep{kay19}. In contrast to more sophisticated MHD models, the ForeCAT model does not account for some time-varying effects, such as magnetic reconnection, and includes the minimal physics necessary to accurately reproduce observed CMEs \citep{kay15b}. ForeCAT uses a static model of the solar magnetic field and approximates the draping magnetic field around a CME to determine the components of the Lorentz force -- magnetic tension and magnetic pressure gradient \citep{kay15b}. These components generate the deflections and are result of the background magnetic field of the star, $\Bsw$.

\begin{figure}
\centering
\includegraphics[width=\columnwidth]{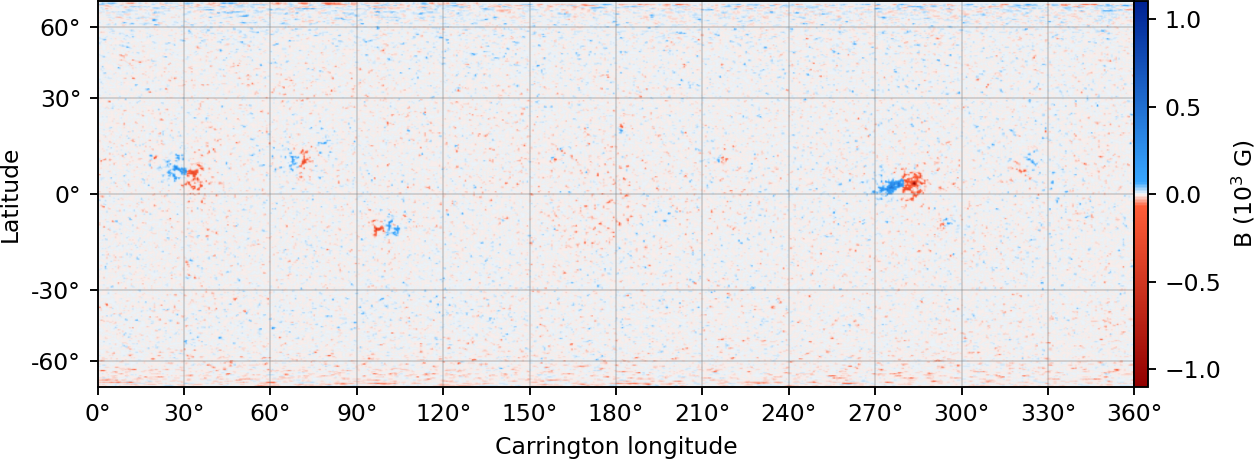}
\caption{Solar photospheric synoptic magnetogram, CR2203, where red represents negative field and blue positive field (in G) on the colour scale.}
\label{fig:CR2203} \end{figure}

Stellar magnetograms are used to determine the harmonic coefficients and construct a Potential-Field Source-Surface model \citep[PFSS;][]{altschuler69, schatten69} of the background magnetic field at a source-surface height (described further). The PFSS model assumes that the magnetic field can be described as current-free below the source-surface and therefore a potential field. For the solar simulations we use a synoptic photospheric magnetograms of the Carrington rotation CR2203 (2018-04-19~05:10 -- 2018-05-16~10:53~UTC, solar minimum, Figure~\ref{fig:CR2203}), generated by the Solar Dynamics Observatory/Helioseismic and Magnetic Imager \citep[SDO/HMI;][]{pesnell12, scherrer12}.

Nonetheless, there are no magnetograms with such resolution for Kepler-63 nor Kepler-411, not even magnetograms reconstructed by the Zeeman-Doppler Imaging technique \citep[ZDI;][]{donati97, vidotto12, vidotto13b, vidotto14b, yu17}. For the stars Kepler-63 and Kepler-411, we extrapolate their magnetograms based on the position and intensity of their starspots obtained from spot transit mapping \citep{netto20, araujo21}. 

\begin{figure}
\centering
\includegraphics[width=.75\columnwidth]{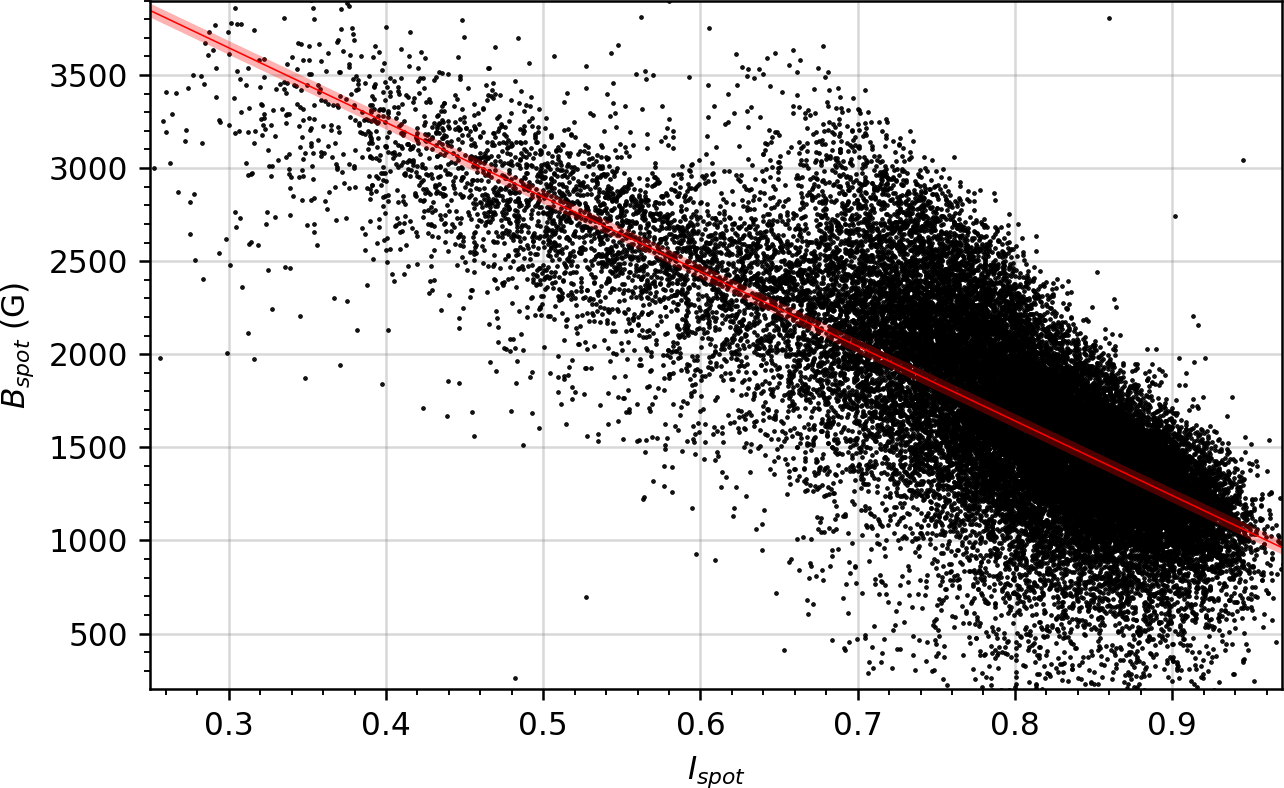}
\caption{Spot's absolute magnetic field, $\Bspot$, as a function of the intensity, $\Ispot$, of sunspots from solar cycle 23. The red line is the linear fit given by Eq.~\ref{eq:I_to_B}}
\label{fig:I_to_B} \end{figure}

The starspots of Kepler-63 and Kepler-411 were mapped by \cite{netto20} and \cite{araujo21}, respectively. \cite{valio03} showed that features on the surface of solar-like stars can be studied using planetary transits. As a planet eclipses its host star, a darker starspot may be obscured, causing a detectable variation in the light curve. With this model (and its improved versions) one may infer the physical properties of the spots, such as size, intensity, position, and temperature \citep{valio17, zaleski19, netto20, zaleski20, selhorst20, araujo21, zaleski22}. To infer the magnetic field intensity of the starspots, we consider the measured intensity of the spots and use a correlation of sunspot intensity and magnetic field. \cite{valio20} analysed physical characteristics of 32,223 sunspots from solar cycle 23, using data from SOHO/MDI \citep{scherrer95}, and established relations between their properties such as area, intensity, temperature, and magnetic field. They presented 4 linear fits of magnetic field, $B$, as a function of spot intensity: positive extreme $B$ field for warm and cold spots, and negative extreme $B$ field for warm and cold spots. In our work, we simplify this relation to a single equation, considering all sunspots and their absolute value of the magnetic field, $\Bspot$ (in G):
\begin{equation} \label{eq:I_to_B}
\Bspot = (4848 \pm 15) - (4008 \pm 20) \times \Ispot\;,
\end{equation}
where $\Ispot$ is the spot intensity, relative to the mean intensity of the centre of the stellar disk, $\Fsct$. The magnetic field of sunspots, $\Bspot$, as a function of $\Ispot$ is shown in Figure~\ref{fig:I_to_B}, where the red line is the linear fit given by Eq.~\ref{eq:I_to_B}. It is worth mentioning that the proposed linear fit is close to the cold spots linear fit by \cite{valio20}, and the group of points suggesting a steeper distribution (at higher intensities) consist of the warm spots.

Here, we assume the linear relation (Eq.~\ref{eq:I_to_B}) to be valid for solar-type G and K stars. Kepler-63 and Kepler-411 have ages of around 200~Myr and may have a much stronger magnetic field than the Sun. However, the linear extrapolations resulted in magnetic fields that we consider to be in agreement with a rotation greater than that of the Sun (Table~\ref{tab:spots}). Then, from the position, intensity and size of the starspots of Kepler-63 and Kepler-411, we were able to reconstruct magnetograms for these stars.

As Kepler-411 has 3 transiting planets, we need to reconstruct maps combining spots probed by different planetary transits \citep{araujo21}. Initially, we set a limit to the time period, $t_\mathrm{lim}=3.0$~days, as the minimum value between the planets orbital period and Kepler-411's rotation period. Then, we combined the transits that were separated in time by less than this time period limit, that is, $|t_i~-~t_{i+ n}|~<~t_\mathrm{lim}$, where $t_i$ is the initial time of a transit and $t_{i+ n}$ is the initial time of the next one. Hence, 282 maps are reconstructed for Kepler-411, out of 196 transits of Kepler-411~b, 76 transits of Kepler-411~c, and 10 transits of Kepler-411~d. The area of the stellar disk that can be probed with these transits are respectively 1.2\%, 2.0\%, and 1.7\%. For Kepler-63, this procedure was not necessary as the spots were probed with transits of a single planet. Therefore, the number of maps was 150, the same number of transits with detected spots, covering 2.8\% of the stellar disk.

\begin{figure}
\centering
\includegraphics[width=\columnwidth]{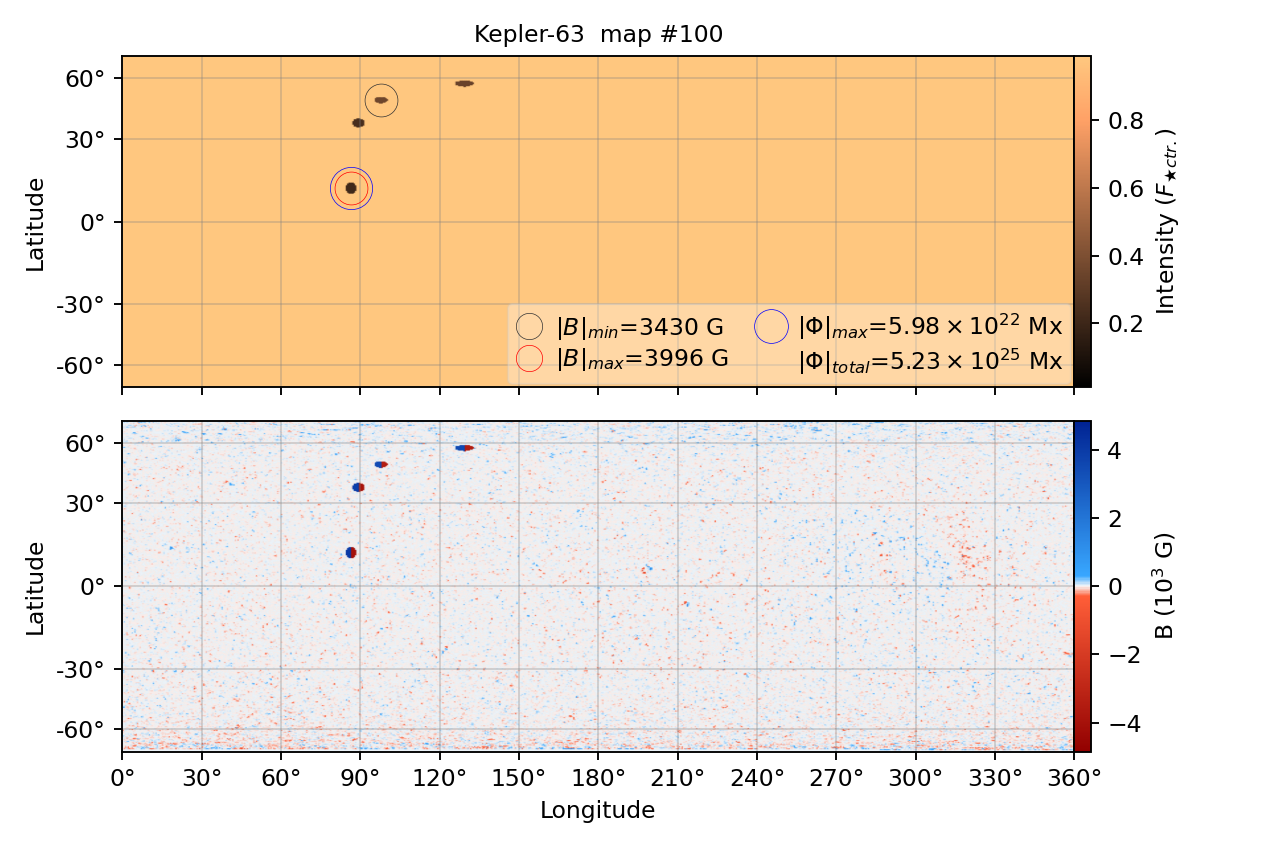} \\[10pt]
\includegraphics[width=\columnwidth]{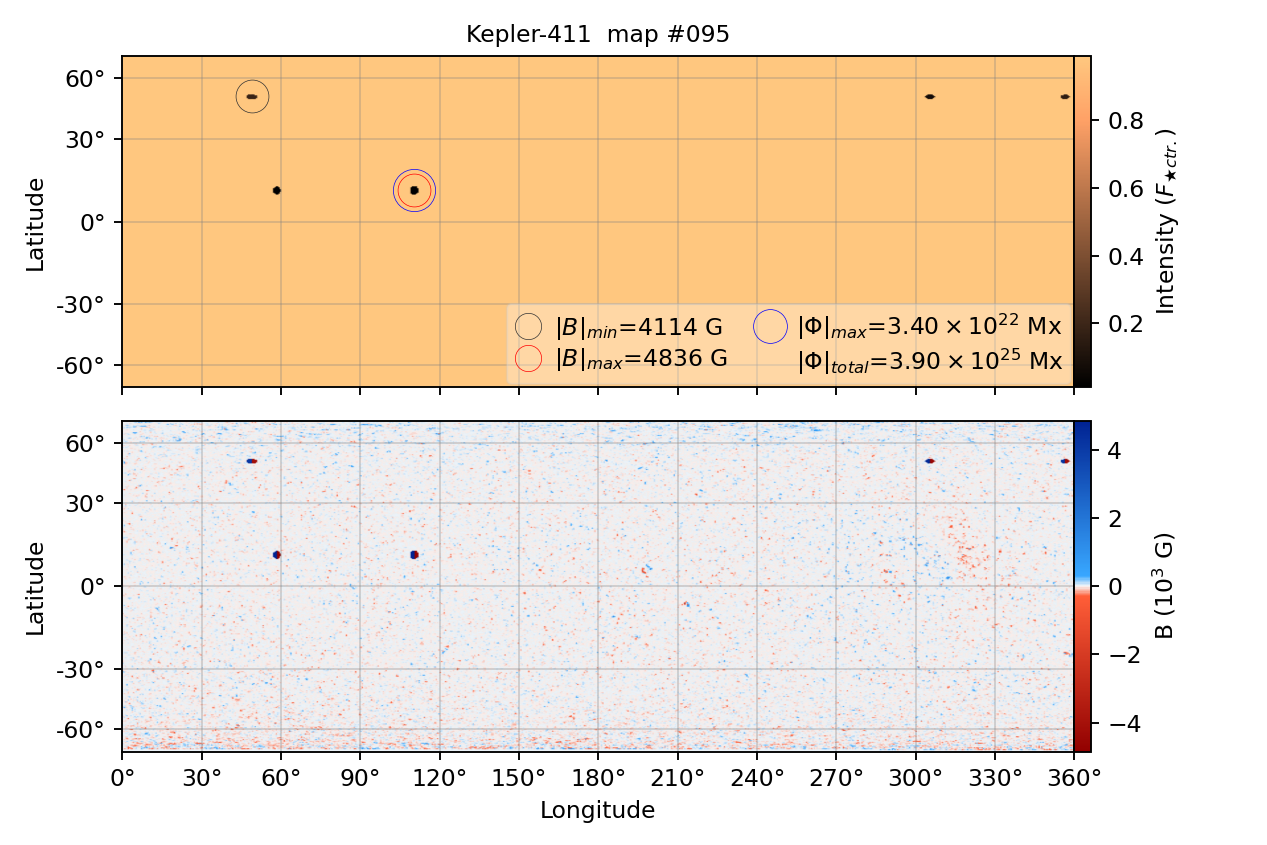}
\caption{Maps of intensity (a, c) and reconstructed magnetograms (b, d) for Kepler-63 (top panels) and Kepler-411 (bottom panels). The intensity is relative by the flux of the stellar disk centre, $\Ispot/\Fsct$, indicated by the orange/brown colour scale. The magnetic field, $B$ (in G), is indicated by the colour scale, where red represents a negative field and blue represents a positive field. The grey circle indicates the spot with the smallest $\Bspot$ and the red circle, the largest $\Bspot$. The blue circle indicates the spot with the highest $\Fspot$. The label $|\Phi|_\mathrm{total}$ corresponds to the magnetic flux absolute values of the whole stellar surface. A video that shows the stellar intensity (spots), magnetograms, background magnetic field, and magnetic field lines can be watched by accessing \href{https://www.youtube.com/watch?v=dgdSCHYdDg8}{this link}.}
\label{fig:I_B_Kepler} \end{figure}

For both models, we then determine the absolute magnetic field, $\Bspot$, for all spots using Eq.~\ref{eq:I_to_B}. We searched for the most extreme maps of each star model, in order to analyse its influence on the CME trajectory results, and compare them with a typical solar case. For this, we categorised the maps into 3 groups: (\textit{i}) the 40 maps with the highest $\Bspot$, (\textit{ii}) the 40 maps with the highest average $\Bspot$, (\textit{iii}) the 40 maps with the highest average magnetic flux. The magnetic flux of a spot is calculated as $\Fspot~=~\Bspot~\times~\Aspot$ , where $\Aspot$ is the spot area. Then, we selected the maps that appeared in the 3 groups. Finally, we select the map with the highest sum of $\Fspot$. The selected map for Kepler-411 corresponds to the \#95 transit combination and for Kepler-63 corresponds to the \#100 transit -- a period of minima during the activity cycle reported by \cite{estrela16}. In Table~\ref{tab:spots} the position and physical parameters of each spot in the maps of both models are listed (latitude, longitude, radius, intensity, magnetic field, and magnetic flux). The intensity maps can be seen in the panels \textit{a} and \textit{c} of Figure~\ref{fig:I_B_Kepler}.

\begin{table}
\centering
\caption{Position and physical parameters for each spot on Kepler-63 and Kepler-411 magnetograms, and for active regions on the Sun for comparison.}
\begin{tabular}{lccccc}
\hline
\mc{6}{c}{Sun (CR2203)}\\
\hline \hline
NOAA AR* (\#)          &  12709 &  12708 &  12706 \\
lat $(^\circ)$         &    6.9 & --10.4 &    2.7 \\
lon $(^\circ)$         &   31.2 &   97.3 &  278.9 \\
$B_\mathrm{avg}$ (G)   &  169.6 &  171.9 &  176.0 \\
\hline
\mc{6}{c}{Kepler-63 (map \#100)}\\
\hline \hline
Spot (\#)              &      1 &      2 &      3 &      4 \\
$\LAspt\;(^\circ)$     &   56.9 &   47.6 &   36.9 &   11.8 \\
$\LOspt\;(^\circ)$     &  129.3 &   97.8 &   89.3 &   86.5 \\
$\Rspot$ (Mm)          &   21.0 &   18.6 &   20.7 &   21.8 \\
$\Ispot\;(\Fsct)$      &  0.339 &  0.354 &  0.239 &  0.213 \\
$\Bspot$ (G)           &   3489 &   3430 &   3890 &   3996 \\
$\Fspot\;(10^{22}$ Mx) &    4.8 &    3.7 &    5.2 &    6.0 \\
\hline
\mc{6}{c}{Kepler-411 (map \#95)}\\
\hline \hline
Spot (\#)              &     1 &     2 &       3 &       4 &      5 \\
$\LAspt\;(^\circ)$     &  49.3 &  49.3 &    49.3 &    11.1 &   11.1 \\
$\LOspt\;(^\circ)$     & 305.5 & 356.6 &    48.9 &    58.4 &  110.4 \\
$\Rspot$ (Mm)          &  11.5 &  11.1 &    12.9 &    13.1 &   15.0 \\
$\Ispot\;(\Fsct)$      & 0.073 & 0.154 &   0.183 &   0.014 &  0.003 \\
$\Bspot$ (G)           &  4556 &  4232 &    4114 &    4793 &   4836 \\
$\Fspot\;(10^{22}$ Mx) &   1.9 &   1.6 &     2.1 &     2.6 &    3.4 \\
\hline
\end{tabular} \label{tab:spots}
{\footnotesize *Active region identification by the National Oceanic and Atmospheric Administration.}
\end{table}

After the selection of maps, the regions of concentrated $B$ field are transformed into simple horizontal dipoles. That is, the areas of the spots with $\Bspot$ distribution have their left half inverted (negative polarity). The rest of the map was filled with the CR2220 magnetogram, a period of solar minimum (2019-07-26~16:04 -- 2019-08-22~21:32~UTC) with low photospheric $B$ field intensity and same magnetic orientation as our solar CME simulations. This magnetogram was multiplied by a factor according to the mean $B$ field strength of the sunspots of each model, taking 1000 G as the mean value of the sunspots $B$ field. 
{That is, the mean B fields of the spots are ~3700 G and ~4500 G, respectively for Kepler-63 and Kepler-411 cases, so the multiplication factors are 3.7 and 4.5}. It is worth mentioning that due to the transits latitudes there are spots mapped in only one hemisphere of Kepler-63 and Kepler-411, which does not correspond to what occurs on the Sun, hence the reconstructed maps are not representative of the true magnetic fields of these stars.

Although the dynamo in relation to age is not linearly proportional, we consider this a reasonable hypothesis. The reconstructed magnetograms yield magnetic flux values of the whole stellar surfaces, $|\Phi|_\mathrm{total}$, of $5.23\times10^{25}$ and $3.90\times10^{25}$~Mx for Kepler-63 and Kepler-411, respectively. \cite{coffaro22} measured logLX~[erg/s]~=~$\sim$29 for Kepler-63, which would correspond to a $|\Phi|_\mathrm{total}$ of about $10^{25}$~Mx, using the relation between average surface magnetic flux of a star and its X-ray luminosity presented by \cite{pevtsov03}, and is close to the $|\Phi|_\mathrm{total}$ we reconstruct in our work. The reconstructed magnetograms can be seen in the panels \textit{b} and \textit{d} of Figure~\ref{fig:I_B_Kepler}.

Using the stellar magnetograms, it is possible to calculate the magnetic forces fields generated by $\Bsw$, with the PFSS model. As mentioned earlier, this model uses a \textbf{source-surface height}, $\Rss$, to calculate $\Bsw$. In solar cases, a default $\Rss$ of 2.5~$R_\odot$ is assumed, however the appropriate $\Rss$ for other stars is unknown. \cite{vidotto11} and \cite{vidotto14} used values of 5 or 4~$R_\star$, respectively, for M dwarf stars, whereas \cite{lang14} used $\Rss~=~2.5~R_\star$.

To determine which $\Rss$ is appropriate, we perform a low-velocity, high-latitude simulation (most extreme cases of deflection), for each $\Rss$ of 2.5, 2.8, and 3.2~$R_\star$. These values were selected because 2.8~$R_\star$ in Kepler-63 case and 3.2~$R_\star$ in Kepler-411 case, are close to 2.5~$R_\odot$ ($\sim$1740~Mm). The results of the simulations showed differences of less than \ang{2} in deflection and rotation of the CMEs of both Kepler-63 and Kepler-411. Therefore, we adopted $\Rss~=~2.5~R_\star$ in all simulations. Finally, we generate the maps of $\Bsw$ at 2.5~$R_\star$ and their respective deflection force fields, with the PFSS model. $\Bsw$ maps, showed in Figure~\ref{fig:CMEs_on_PFSS}, tend to approach the Astrospheric Current Sheet shape, and have higher intensities located at the poles due to coronal holes. A video that shows the stellar intensity (spots), the magnetogram, the background magnetic field (calculated with PFSS), and magnetic field lines (calculated with PFSS) for each star model, can be watched by accessing \href{https://www.youtube.com/watch?v=dgdSCHYdDg8}{this link}.
\subsection{CME Parameters} \label{sec:model_CME}

We set different CME's initial positions on the star (\textbf{latitude}, $\lati$, and \textbf{longitude}, $\loni$) for each simulation. CME's \textbf{inclination} is measured clockwise with respect to the equatorial plane, where $\inci=\ang{0}$. The shape and size of the CME are also set. The model simulates the behaviour of a 3D rigid half torus, similar to other models \citep{gibson98, titov99, chen96, thernisien06}, intended to represent the flux rope of a CME as a function of radial distance \citep{kay15b}. CMEs can evolve during propagation, undergoing some effects such as "pancaking", erosion, or other distortions \citep{savani11, riley04, riley04b, nieves18}. However, according to observations and magnetohydrodynamic (MHD) simulations they maintain the flux rope structure reasonably up to 1~AU, remaining predominantly in a torus shape \citep{burlaga81, klein82, cane03, vandas02}.

In the model, the toroidal axis describes a half ellipse -- \textit{i.e.} half of the CME -- and the torus has a circular cross section \citep{kay15b}. The torus surface, $X$, is defined in terms of the toroidal and poloidal angles, $\theta_t$ and $\theta_p$:
\begin{equation} \label{eq:torus}
\begin{split}
X(\theta_t,\theta_p) = 
[ \; (a+b\cos{\theta_p})\cos{\theta_t}\;,\; b\sin{\theta_p}\;,\\ (c+b\cos{\theta_p})\sin{\theta_t} 
\; ] \;,
\end{split}
\end{equation}
where $a$, $b$ and $c$ are the \textbf{shape parameters}. In the toroidal direction $a$ is the axis towards the CME nose, and $c$ is the perpendicular axis. The cross section is described by the radius $b$. For our simulations, we set the \textbf{ratios} $a/c~=~1$ and $b/c~=~0.2$. The torus size increases in the same rate as the angular width, $\AWr$ (described next), with \textbf{initial cross section radius}, $b_\circ~=~0.02$. Moreover, the surface is a \textbf{grid} of $\Npol\times\Ntor$ points in the poloidal and toroidal directions, respectively. According to \cite{kay15b}, the solution converges to grids up to $13~\times~15$, which are the values we use.

There are models and observational data that show different CME velocity profiles, \textit{e.g.} \cite{sheeley99}: some fast solar CMEs -- usually connected with strong flares -- can be accelerated up to several 1000 km/s within a few solar radii, then decelerate to the solar wind speed. For our simulations, we provided the model (same used by \citep{kay13, kay15b, kay16}) with empirical models for angular width and velocity as a function of radial distance, $r$, that rapidly increases in the low corona then remains constant beyond about 5 stellar radii, which is consistent with solar observations \citep{aschwanden09, patsourakos10a, patsourakos10b}. More precisely, the CME's angular width is defined as:
\begin{equation} \label{eq:AW}
\AWr = \AWi + (\AWf- \AWi) \left(1 - e^{-(r-1)/r_\omega}\right)\;,
\end{equation}
where the \textbf{initial} and \textbf{final angular widths} are respectively $\AWi~=~\ang{8}$ and $\AWf~=~\ang{44}$, and the \textbf{length scale} over which the width varies is $r_\omega~=~1.5$. The CME trajectories are simulated from $r_\circ~=~1.1~R_\star$ to $r_f~=~160~R_\star$, since after this point the deflections are very low or null, keeping the trajectories stable.

\begin{figure}
\centering
\includegraphics[height=.7\columnwidth]{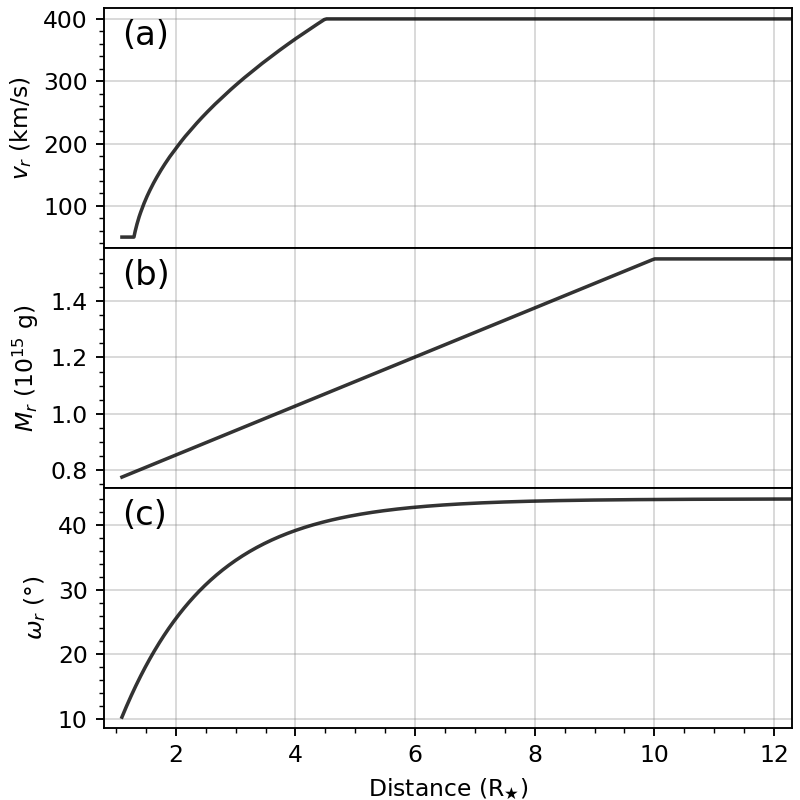}
\caption{Velocity, $\vr$ (a), mass, $M_r$ (b), and angular width, $\AWr$ (c), as a function of radial distance, $r$, between 1 and 12~$R_\star$.}
\label{fig:FCATmodels} 
\end{figure}

Another aspect to consider on the CME's propagation is the mass. \cite{vourlidas10} analysed the mass evolution of CMEs in the lower corona and found that CMEs tend to increase in mass in the solar corona up to 10~$R_\odot$. From this point on, the mass is considered constant. The model used along the propagation determines the evolution of the CME mass, $M_r$, as
\begin{equation} \label{eq:M}
\begin{split}
M_r = \frac{\Mcme}{2} \left( 1 + \frac{r - r_\circ}{\rM - r_\circ} \right) \hspace{3.5ex} \for r \leq \rM\;,\\
M_r = \Mcme \hspace{18ex} \for r > \rM\;, 
\end{split}
\end{equation}
where $\Mcme$ is the final mass of the CME, and the radial position that limits the increase of $M_r$ is $\rM~=~10~R_\star$, keeping it constant from this point forward. \cite{vourlidas10} analyzed 7668 CMEs using data from the {Solar \& Heliospheric Observatory} \citep[SOHO;][]{domingo95}, and found typical CME masses between $10^{12}$~g and $10^{16}$~g, and calculated an average mass of $1.55~\times~10^{15}$~g, which is the value we adopt for $\Mcme$ in the simulations. Figure~\ref{fig:FCATmodels} shows the variation of the parameters $\vr$, $M_r$, and $\AWr$ as a function of $r$, between 1 and 12~$R_\star$.

For the CME's velocity, we use a three-phase propagation model, similar to that presented by \cite{zhang06}: a gradual rise followed by an impulsive acceleration and, finally, a constant radial propagation. In this model, the three propagation phases respectively characterise the radial velocities, $\vr$, of a CME:
\begin{equation} \label{eq:Vr}
\begin{split}
\vr = v_\circ \hspace{26.4ex} \for R_\odot \leq r \leq \rGA\;,\:\\
\vr = \sqrt{v_\circ^2 + \frac{(v_f^2 - v_\circ^2)(r - \rGA)}{(\rAP - \rGA)}} \hspace{4ex} \for \rGA \leq r \leq \rAP\;,\,\\
\vr = v_f \hspace{25.9ex} \for \rAP < r\;,\hspace{6.4ex} 
\end{split}
\end{equation}
where $v_\circ$ and $v_f$ are the \textbf{initial} and \textbf{final radial velocities}, and the fixed \textbf{radial positions} to transition between the three phases are $\rGA~=~1.3~R_\star$ and $\rAP~=~4.5~R_\star$. We use 3 velocity configurations: \textbf{low}-velocity with $v_\circ~=~50$~km/s and $v_f~=~400$~km/s, \textbf{mid}-velocity with $v_\circ~=~70$~km/s and $v_f~=~650$~km/s, and \textbf{high}-velocity with $v_\circ~=~90$~km/s and $v_f~=~900$~km/s.

The \textbf{initial position} ($\loni$, $\lati$, $\inci$) of the CME is based on regions of concentrated $B$ field, lining up the torus with polarity inversion lines, locations where flux ropes would be likely to form. We set 3 different values of $\lati$ for each model, so that there are \textbf{low}-, \textbf{mid}- and \textbf{high}-latitude configuration. In Table~\ref{tab:posicao0}, we list the starting position of CMEs for all simulations. During CR2203, there were some low-intensity solar events related to partial halo CMEs\footnote{SOHO CELIAS Proton Monitor, Solar Wind Parameters for Carrington Rotation 2203 (hour averages), \url{https://space.umd.edu/pm/crn/crn_2203.html}}. However, we place a hypothetical CME in the active region of longitude \ang{70} and latitude \ang{10} (low latitude). In the mid- and high-latitude simulations, we kept the same longitude and varied the latitudes respectively for \ang{30} and \ang{45}. For Kepler-63 simulations, we place the CMEs in the starspot regions with the exception of the highest latitude spot. For Kepler-411, the low latitude CME is placed in the spot with the highest $\Bspot$ ($\lati=\ang{11.1}, \loni=\ang{110.4}$), and the high latitude CME is placed in the spot of $\lati=\ang{49.3}$ and $\loni=\ang{48.9}$. The mid-latitude CME was positioned between the two previous ones, as there is no mid-latitude spot on this map.

\begin{table}
\centering
\caption{Initial position of CMEs for each star model and latitudinal region.}
\begin{tabular}{lccc}
\hline
Star   & Latitude & \mc{2}{c}{Initial position} \\
         & region & $\lati$    & $\loni$    \\
\hline \hline
Sun        & high & \ang{45,0} & \ang{70,0} \\
           & mid  & \ang{30,0} & \ang{70,0} \\
           & low  & \ang{10,0} & \ang{70,0} \\
\hline
Kepler-63  & high & \ang{47,6} & \ang{97,8} \\
           & mid  & \ang{36,9} & \ang{89,3} \\
           & low  & \ang{11,8} & \ang{86,5} \\
\hline
Kepler-411 & high & \ang{49,3} & \ang{48,9} \\
           & mid  & \ang{30,0} & \ang{80,0} \\
           & low  & \ang{11,1} & \ang{110,4} \\
\hline
\end{tabular} \label{tab:posicao0} \end{table}
\subsection{Stellar Parameters} \label{sec:model_stellar}

The model also includes the non-radial drag effects resulting from the interaction between a CME and the solar wind. To calculate the volumetric force due to drag, \cite{kay13, kay15b} adapted the expression of \cite{cargill96} and \cite{cargill04} for the non-radial direction. The \textbf{drag coefficient adopted} is $\CD~=~1$. 

Despite stellar age playing a key role in the mass loss of stars, we adopted the same solar wind density model for all simulations, because drag effects on CMEs are very weak compared to the magnetic forces in our simulations. To determine the background solar wind density, the model uses the \cite{guhatha06} model that empirically scales the value based on the distance from the Astrospheric Current Sheet and the radial distance. The radial solar wind speed can be calculate by assuming constant mass flux.

Moreover, the stellar \textbf{radius}, $R_\star$, and \textbf{rotation rate}, $\Omega_\star$, must be set. For the Sun the values are respectively $7.0\times10^{10}$~cm and $2.8\times10^{-6}$~rad/s; for Kepler-63, $6.3\times10^{10}$~cm and $1.3\times10^{-5}$~rad/s \citep{sanchis13, netto20}; and for Kepler-411, $5.5\times10^{10}$~cm and $6.9\times10^{-6}$~rad/s \citep{araujo21}.
\section{Deflections on the CMEs' Trajectories} \label{sec:results}

\begin{figure*}
\centering
\includegraphics[width=.8\textwidth]{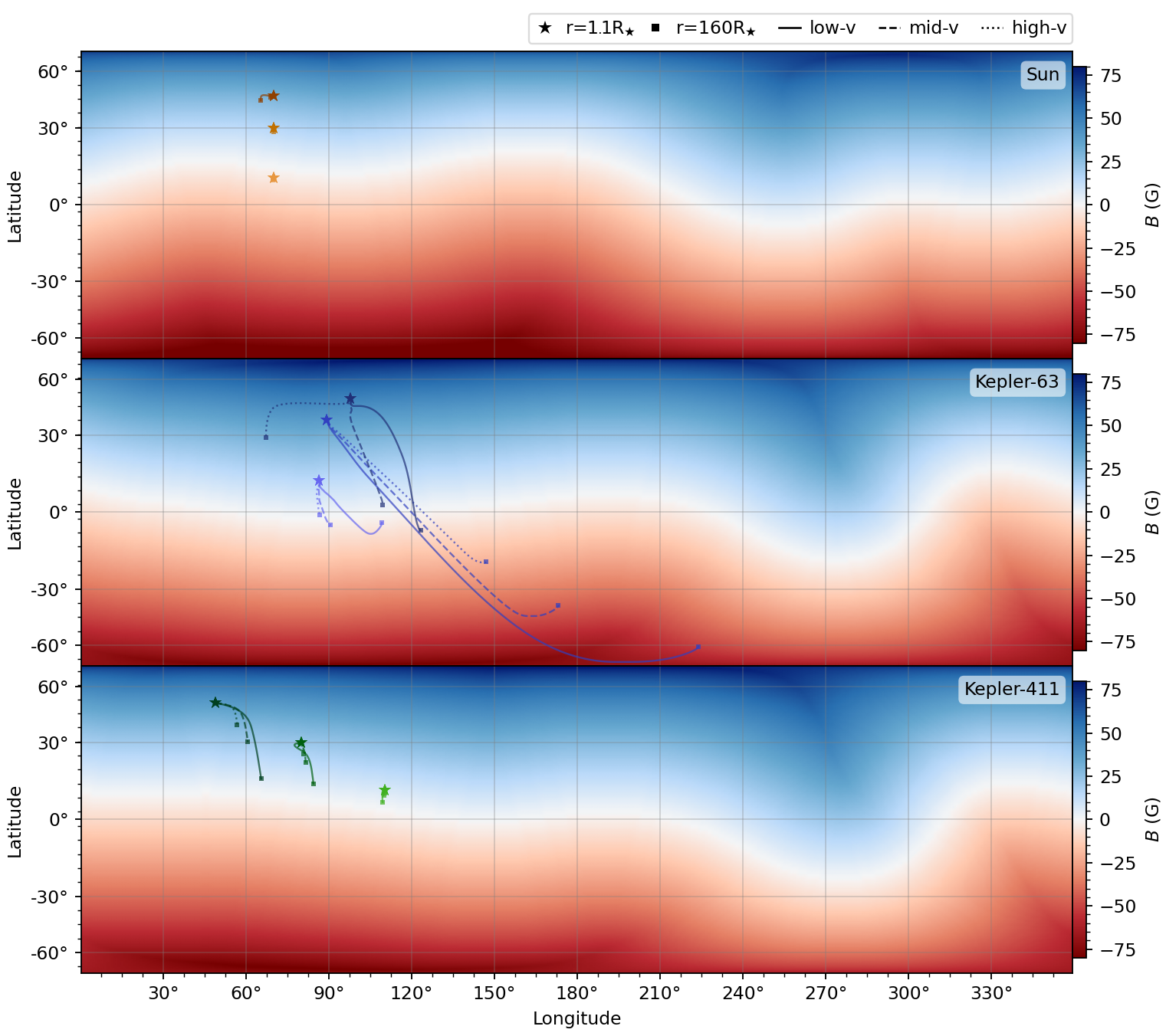}
\caption{Background magnetic field, $\Bsw$, maps for the Sun (top panel), Kepler-63 (middle panel), and Kepler-411 (bottom panel), calculated with the PFSS model using $\Rss~=~2.5~R_\star$. colour scale represents $\Bsw$ in G. CME's trajectories are over-plotted on the $\Bsw$ maps, and the 3 cases (latitudes) are shown for each star. The star symbol represents the initial position ($r~=~1.1~R_\star$) and the square represents the final position ($r~=~160~R_\star$) of the CMEs. The solid, dashed and dotted lines represent respectively low-, mid- and high-velocity simulations.}
\label{fig:CMEs_on_PFSS} 
\end{figure*}

The variations in CME position (latitude, longitude), rotation, and velocities (latitudinal, longitudinal) are shown in Figures~\ref{fig:SUN_l1}, \ref{fig:SUN_l2}, \ref{fig:SUN_l3}, \ref{fig:K63_l1}, \ref{fig:K63_l2}, \ref{fig:K63_l3}, \ref{fig:K411_l1}, \ref{fig:K411_l2}, and \ref{fig:K411_l3}, in Appendix~\ref{sec:outFCAT}, for each model and for each initial latitude launch. These parameters are showed as a function of radial distance, from 1.1 to $\sim$45~$R_\star$ to emphasise the deflections close to the star. The stellar rotation rate, $\Omega_\star$, is subtracted from longitude variation, so the referential longitude axis follows the stellar rotation. In addition, we made videos which show the evolution of the 3D trajectory: latitude, longitude (considering stellar rotation), rotation, and geometry (shape, size and expansion) of the CMEs simulated by the ForeCAT model. The videos are available as supplementary material and show the simulations from 1.1 to $\sim$45~$R_\star$. Moreover, we over-plotted the complete CME's trajectories on the $\Bsw$ maps, which are showed in Figure~\ref{fig:CMEs_on_PFSS}.

We calculate the total variations of longitude, $\Dlon$, latitude, $\Dlat$, and rotation, $\Drot$, over complete trajectories ($1.1-160~R_\star$) -- the value at final position minus value at the initial position. In Table~\ref{tab:DeflTot}, the total variations are listed by model, initial latitude, $\lati$, and velocities, $v_\circ$ and $v_f$. In general, the most intense longitude deflections occur up to $\sim$2~$R_\star$, which can be noticed by the high longitude gradient and by the peaks and valleys of longitudinal velocity, $\vlon$, in Figures~\ref{fig:SUN_l1}, \ref{fig:SUN_l2}, \ref{fig:SUN_l3}, \ref{fig:K63_l1}, \ref{fig:K63_l2}, \ref{fig:K63_l3}, \ref{fig:K411_l1}, \ref{fig:K411_l2}, and \ref{fig:K411_l3}. Following the same trend, the rotation has more intense variations near the surface of the stars.

\begin{table}
\centering
\caption{Total variations of longitude, $\Dlon$, latitude, $\Dlat$, and rotation, $\Drot$, over the simulated CME trajectories.}
\begin{tabular}{lcccccc}
\hline \\[-7pt] 
Star & $\lati$ & $v_\circ$ & $v_f$ & $\Dlon$ & $\Dlat$ & $\Drot$ \\
 & ($^\circ$) & (km/s) & (km/s) & ($^\circ$) & ($^\circ$) & ($^\circ$) \\[1pt]
\hline \hline \\[-7pt] 
Sun         & 10   & 50 & 400 &  --0.1 &  --0.8 &   --3.9 \\
(low-lat.)  & 10   & 70 & 650 &  --0.1 &  --0.4 &   --1.9 \\
            & 10   & 90 & 900 &  --0.1 &  --0.2 &   --1.1 \\[4pt]
Sun         & 30   & 50 & 400 &  --0.0 &  --1.6 &     0.1 \\
(mid-lat.)  & 30   & 70 & 650 &  --0.0 &  --0.7 &     0.0 \\
            & 30   & 90 & 900 &  --0.0 &  --0.4 &     0.0 \\[4pt]
Sun         & 45   & 50 & 400 &  --4.7 &  --2.3 &   --3.8 \\
(high-lat.) & 45   & 70 & 650 &  --1.1 &  --0.7 &   --2.3 \\
            & 45   & 90 & 900 &  --0.5 &  --0.3 &   --1.5 \\
\hline \\[-7pt] 
Kepler-63   & 11.8 & 50 & 400 &   22.4 & --15.9 & --365.4 \\
(low-lat.)  & 11.8 & 70 & 650 &    3.9 & --16.6 & --445.0 \\
            & 11.8 & 90 & 900 &    0.2 & --13.0 & --304.9 \\[4pt]
Kepler-63   & 36.9 & 50 & 400 &  132.1 & --97.9 &   --4.0 \\
(mid-lat.)  & 36.9 & 70 & 650 &   82.9 & --74.1 &    17.5 \\
            & 36.9 & 90 & 900 &   57.5 & --55.7 &    17.5 \\[4pt]
Kepler-63   & 47.6 & 50 & 400 &   25.0 & --54.5 & --165.9 \\
(high-lat.) & 47.6 & 70 & 650 &   11.0 & --45.0 &  --31.1 \\
            & 47.6 & 90 & 900 & --30.6 & --18.7 &  --17.6 \\
\hline \\[-7pt] 
Kepler-411  & 11.1 & 50 & 400 &  --0.9 &  --4.6 &   --0.1 \\
(low-lat.)  & 11.1 & 70 & 650 &  --0.3 &  --2.2 &   --0.0 \\
            & 11.1 & 90 & 900 &  --0.2 &  --1.2 &   --0.0 \\[4pt]
Kepler-411  & 30   & 50 & 400 &    4.4 & --16.5 &   --2.2 \\
(mid-lat.)  & 30   & 70 & 650 &    1.6 &  --8.3 &   --1.9 \\
            & 30   & 90 & 900 &    0.8 &  --4.8 &   --1.3 \\[4pt]
Kepler-411  & 49.3 & 50 & 400 &   16.4 & --34.0 &   --7.8 \\
(high-lat.) & 49.3 & 70 & 650 &   11.6 & --18.8 &   --6.7 \\
            & 49.3 & 90 & 900 &    7.9 & --11.1 &    -4.9 \\
\hline
\end{tabular} \label{tab:DeflTot} \end{table}

The latitude deflections are more intense in the solar model for low-latitude up to $\sim$3~$R_\star$ in the Kepler-63 models for low- and mid-latitude up to $\sim$2~$R_\star$. For the other simulations, the latitude deflections are more pronounced up to $\sim$6~$R_\star$, and after this point they still occur, but with lower values. This is largely due to the influence of concentrated $B$ field regions (spots and/or active regions), which cause the CME trajectory to deflect and rotate closer to the photosphere. On the other hand, latitude varies gradually along the propagation, as $\Bsw$ tends to approach Astrospheric Current Sheet from the source-surface outwards, and CMEs tend to deviate to regions of minimal magnetic energy \citep{cremades04, kilpua09, gopalswamy09, kay15b}. In low-latitude simulations for the Sun, although more intense up to $\sim$2~$R_\star$, the absolute values of total latitude variations, $|\Dlat|$, are very low, smaller than \ang{1} (Figure~\ref{fig:CMEs_on_PFSS} and Table~\ref{tab:DeflTot}). In the Kepler-63 case, $|\Dlat|$ is more significant ($\sim$\ang{13}--\ang{17}) for low-latitude simulations, and more extreme ($\sim$\ang{56}--\ang{95}) for mid-latitude simulations.
\section{Discussion and Conclusions} \label{sec:conclusion}

We used the ForeCAT model to perform a total of 27 simulations of CME trajectories for the Sun, Kepler-63, and Kepler-411. The input parameters of the CMEs (Section~\ref{sec:model_CME}) were kept the same for all simulations, with the exception of initial positions ($\loni$ and $\lati$, Table~\ref{tab:posicao0}), velocities ($v_\circ$ and $v_f$), stellar parameters ($R_\star$ and $\Omega_\star$), and background magnetic fields, $\Bsw$. For calculation of the $\Bsw$ maps for Kepler-63 and Kepler-411, we reconstructed magnetograms by extrapolating the $B$ field of their respective mapped starspots \citep{valio20, netto20, araujo21}. Furthermore, trajectories were simulated from $r_\circ~=~1.1~R_\star$ to $r_f~=~160~R_\star$.

All CMEs show latitude variations toward the Astrospheric Current Sheet, except for Kepler-63 mid-latitude simulations. CMEs tend to follow this trend on global scales, although the magnetic structure in the lower corona can also affect the direction of deflection and rotation. This is because the magnetic forces responsible for the deflection of CMEs decay quickly over radial distance, which causes most of the deflection and rotation to occur near the Sun \citep{kay13, kay15b}. The magnetic pressure gradient is the force that most contributes to the deflection of CMEs towards the Astrospheric Current Sheet. However, the intense magnetic field of the spots cause greater deflection in the CMEs, so that there are cases where the CME propagates towards high latitudes.

In general, the absolute total variations are small for solar simulations, intermediate for Kepler-411 simulations, and intense in the Kepler-63 cases. This is a trend that follows the photospheric $B$ field of stars: the solar magnetogram has a $B$ field peak just above 1000~G, while the magnetograms of Kepler-63 and Kepler-411 have peaks of 3996 and 4836~G, respectively (Figures~\ref{fig:CR2203} and \ref{fig:I_B_Kepler}, and Table~\ref{tab:spots}). Also, although Kepler-411's spots have larger $\Bspot$, their radii, $\Rspot$, are approximately 2 times smaller than Kepler-63's, thus their $\Fspot$ are about 2 to 3 times larger (Table~\ref{tab:spots}) since $\Fspot\propto\Rspot^2$. In addition, $\Rspot$ being larger in Kepler-63 model, leads to larger magnetic loops formation and a greater area of influence in the deflection and rotation of the CMEs.

The Kepler-63 simulation shows much larger CME deflections and rotation than those of solar CMEs. Kepler-63 has stronger ambient magnetic field as well as stronger concentrated magnetic field in active regions, as compared to the Sun, so it is not unreasonable to find more extreme deflections and rotations, driven by the more extreme magnetic forces. It is also possible that in this intense magnetic environment, CMEs could have higher masses than their solar counterpart. \cite{kay15b} compare trajectories of CMEs with different masses and show that more massive CMEs reach slower non-radial velocities, and thus lower deflections. If this is the case, which we cannot currently say given the lack of observations, then smaller deflections and rotations are expected, more similar to what is seen in the Sun. However, if the average solar CME mass is appropriate, then we would expect the extreme deflections and rotations seen in Kepler-63 simulations.

The absolute total variations of deflections and rotations increased with $\lati$, \textit{i.e.}, these were greater for higher initial latitudes. Part of this is due to the combination of concentrated field regions (strong deflection forces) and the tendency of CMEs to deviate towards the Astrospheric Current Sheet. Moreover, in most cases $|\Dlon|$, $|\Dlat|$, and $|\Drot|$ are inversely proportional to $v_\circ$ and $v_f$. This is mainly because faster CMEs move away more quickly from the lower corona, where the deflections are more significant. However, the Kepler-63 simulations do not follow neither the latitude nor the velocity trends. Considering the 3 models, Kepler-63 has the largest spots with the most intense $\Fspot$. Furthermore, the spatial configuration of active regions is very particular in each model. On the surface of Kepler-63, the spots are distributed very close to each other, which also strongly influences the large variations of $\Dlon$, $\Dlat$, and $\Drot$.

\cite{kay16} reported that planets, with orbits in a plane close to the equatorial plane of the star, are more likely to be hit by CMEs. Kepler-63's CMEs, with mid-latitude and with low- and mid-velocities, were ejected from the northern hemisphere and deflected to the point of heading towards the south pole. In these cases, the probability of the exoplanet Kepler-63~b being hit by CMEs would increase, since it has a near polar orbit \citep{sanchis13, netto20}. Another interesting aspect of this star is a large, long-lived polar spot \citep{sanchis13}. Could this be some kind of CME overproducing region? Considering high ejection latitudes and the fact that high-velocity CMEs tend to have more radial trajectories, Kepler-63~b could become a common target for CME impacts. Also, it is worth mentioning that even if an active star is potentially more likely to produce many CMEs (towards the poles in the case of Kepler-63, or deflected to the Astrospheric Current Sheet in the case of Kepler-411), which would increase the number of potential hits, the strong magnetic field might not allow the CME to erupt due to magnetic confinement \citep{alvarado18, sun22}, thus decreasing CME detection rates.

The highest values of $\Drot$ occur in Kepler-63 simulations; in some cases the CMEs rotate more than \ang{360}. The CMEs rotate around the axis of radial propagation, so $\Drot$ can be an approximation of the direction of the field $B$ resulting from the CME. This is another determining factor in the impact generated on space weather near a planet. The CME $B$ field tends not to generate large changes in the $B$ field of a planet, when they have the same directions, generally causing compression of the planetary $B$ field. However, a CME with field $B$ in the opposite direction can generate large changes in the planetary field $B$, as there may be reconnection between the lines of both fields when they collide. Thus, it can be said that a more homogeneous magnetic field such as the solar one (CR2203) provides CMEs that are more likely to have low deflections, while a magnetogram such as that of Kepler-63 generates less stable trajectories.

Furthermore, as the velocity of CME increases, their trajectories tend to become more radial. \cite{xie09} observed a correlation between the CME deflection and the distance between the CME source and the streamer belt for the slow CMEs, but did not find this correlation for the fast CMEs, which tend to deviate statistically less than the slow ones. Our simulations that best fit this trend are the Kepler-411 cases. Also, solar CMEs can be faster than our models (400~km/s$\geq{v_f}\geq$900~km/s). Such fast velocities could yield different deflections in the trajectories.

CMEs have adverse effects on near-Earth space weather, however the knowledge about stellar CMEs is still incipient. Although we used approximations and hypotheses that are not yet well established, we were able to simulate CME trajectories for the star models Kepler-63 and Kepler-411. Furthermore, we show that the free parameters affect the trajectories of CMEs. In general, deflections and rotations of CMEs decrease with radial velocity, and increase with latitude from the initial position. However, strength and configuration of magnetic field are certainly the main driver of deflections. In the future, new approaches can be explored in this kind of study, like varying more parameters of the CMEs, such as initial inclination, angular width and mass, and mainly, using different magnetic backgrounds for more star models.
\backmatter

\bmhead{Supplementary material} \label{sec:suppl}

Videos of the CME simulations are available at 
\url{https://www.youtube.com/playlist?list=PLX4O1akWIZ3do63Fo9qO7YKbaNDqxnt1l}, where each video can be watched directly, and at 
\url{https://drive.google.com/file/d/1Kq4W_TUd5S7Y6O--P7pHmPy7iCTP26KF/view?usp=sharing}, where the file \texttt{CME\_videos.zip} can be downloaded. 
More information on Appendix \ref{sec:outFCAT}.
\bmhead{Data Availability}

The data from the SDO/HMI underlying this article are available as FITS files in the Joint Science Operations Center (JSOC) page at \url{http://jsoc.stanford.edu/ajax/lookdata.html?ds=hmi.Synoptic_Mr_720s}.
\bmhead{Acknowledgments}

F. M. acknowledges financial support from the Fundo Mackenzie de Pesquisa e Inovação (MackPesquisa), the Coordenação de Aperfeiçoamento de Pessoal de Nível Superior 
(CAPES), the French National Research Agency funded project PEPPER (ANR-20-CE31-0002), and the Fundação de Amparo à Pesquisa do Estado de São Paulo (FAPESP; process numbers 2022/12024-0 and 2013/10559-5). A. A. acknowledges financial support from the Conselho Nacional de Desenvolvimento Científico e Tecnológico (CNPq; \#150817/2022-3). The CR2203 synoptic photospheric magnetograms is a courtesy of NASA/SDO and the AIA, EVE, and HMI science teams.
%
%
%
%
\bibliography{Ref}
\addcontentsline{toc}{section}{References}
%
%
%
%
\vfill 
\begin{appendices}

\section{}
\addcontentsline{toc}{section}{ForeCAT's Output Parameters}
\subsection*{ForeCAT's Output Parameters} \label{sec:outFCAT}

We present the results of the outputs of CME trajectory simulations, in addition to videos of 3D CME trajectories as supplementary material (\href{https://www.youtube.com/playlist?list=PLX4O1akWIZ3do63Fo9qO7YKbaNDqxnt1l}{videos link}). The videos show simulations from 1.1 to $\sim$45~$R_\star$ to focus on the deflections close to the star. In order to identify the results of each simulation more quickly, we use the following colour scheme: orange represents solar simulations, blue represents Kepler-63, and green, Kepler-411. Furthermore, according to the latitude configuration, each of these colours goes from lighter (low latitudes) to darker tones (high latitudes). Moreover, the videos and figures have the indications "lat$_1$", "lat$_2$" and "lat$_3$", according to low-, mid- and high-latitude, respectively. Following the same logic, the indications "v$_1$", "v$_2$" and "v$_3$" are according to low-, mid- and high-velocity configurations, respectively.

\pagebreak

\begin{figure}
\centering
\includegraphics[width=.99\columnwidth]{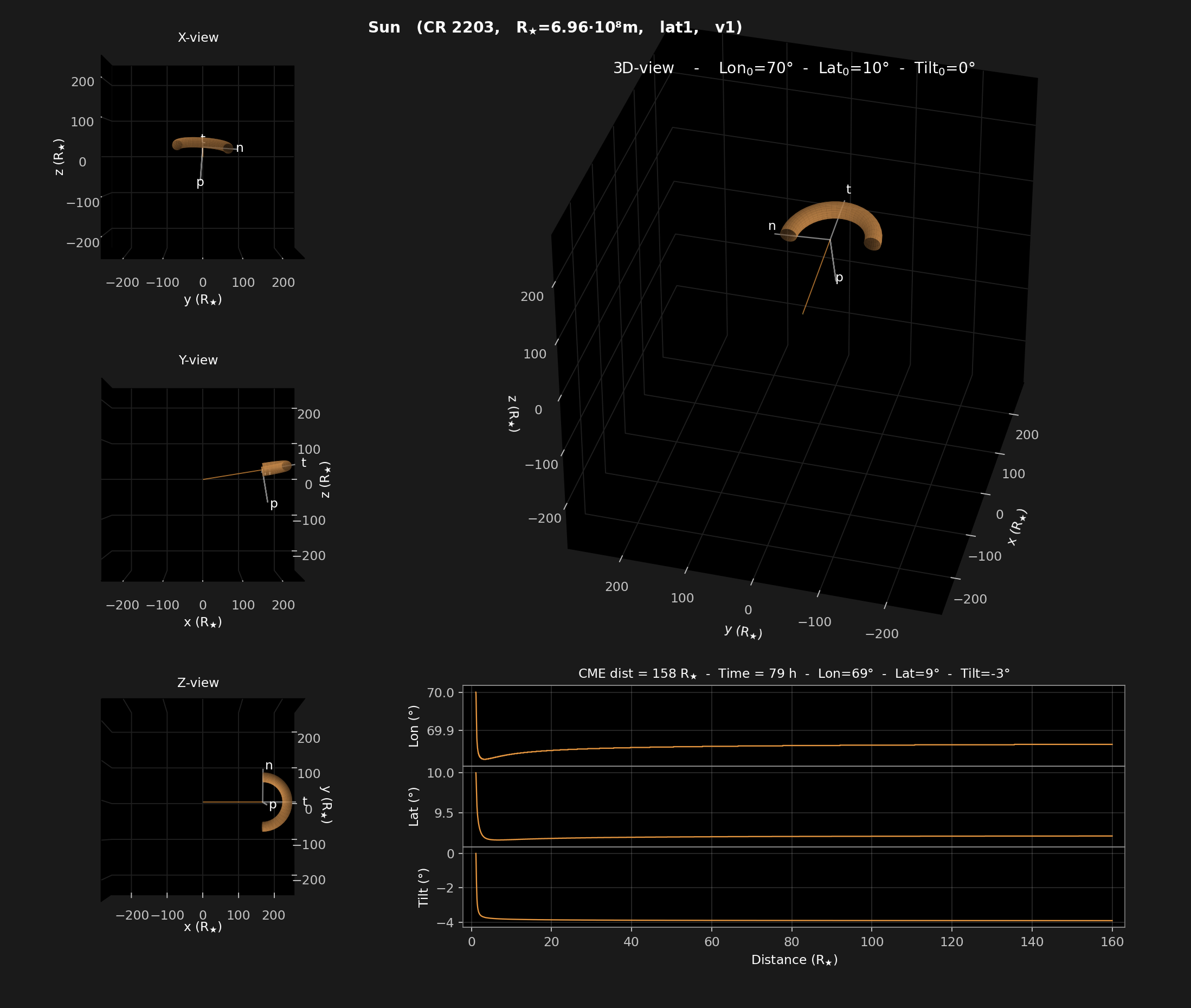}
\caption{3D representation of the final instant ($r=160~R_\star$) of the CME trajectory simulation for the \textbf{Sun} model, \textbf{low-latitude}, \textbf{low-velocity}. The panels on the left respectively correspond to views on the $x$, $y$ and $z$ axis. The upper panel on the right corresponds to a 3D point of view that varies throughout the simulation. The group of lower panels on the right show, respectively, the longitudinal and latitudinal variations, and the rotation of the CME, over its trajectory, up to 160~$R_\star$. The video version 
can be accessed via the links in the Supplementary Material section or \href{https://www.youtube.com/watch?v=Wf3-ueNd56Q}{this link}.}
\label{fig:3D-SUN-lat1-v1} \end{figure}

\begin{figure}
\centering
\includegraphics[width=.99\columnwidth]{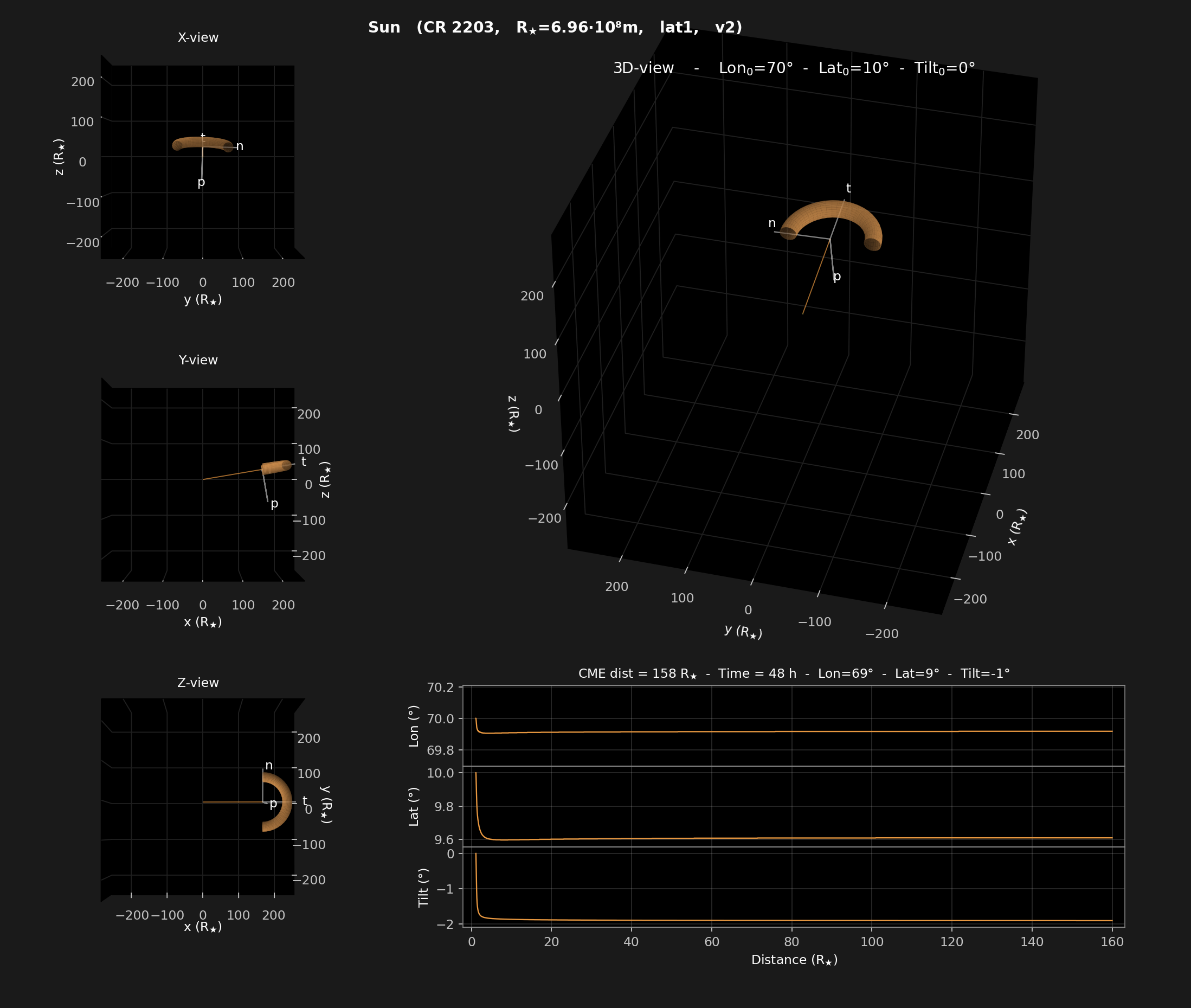}
\caption{Same as Figure~\ref{fig:3D-SUN-lat1-v1}, but for \textbf{Sun} model, \textbf{low-latitude}, \textbf{mid-velocity}. The video version can be accessed via the links in the Supplementary Material section or \href{https://www.youtube.com/watch?v=8-uPG_F30HE}{this link}.}
\label{fig:3D-SUN-lat1-v2} \end{figure}

\begin{figure}
\centering
\includegraphics[width=.99\columnwidth]{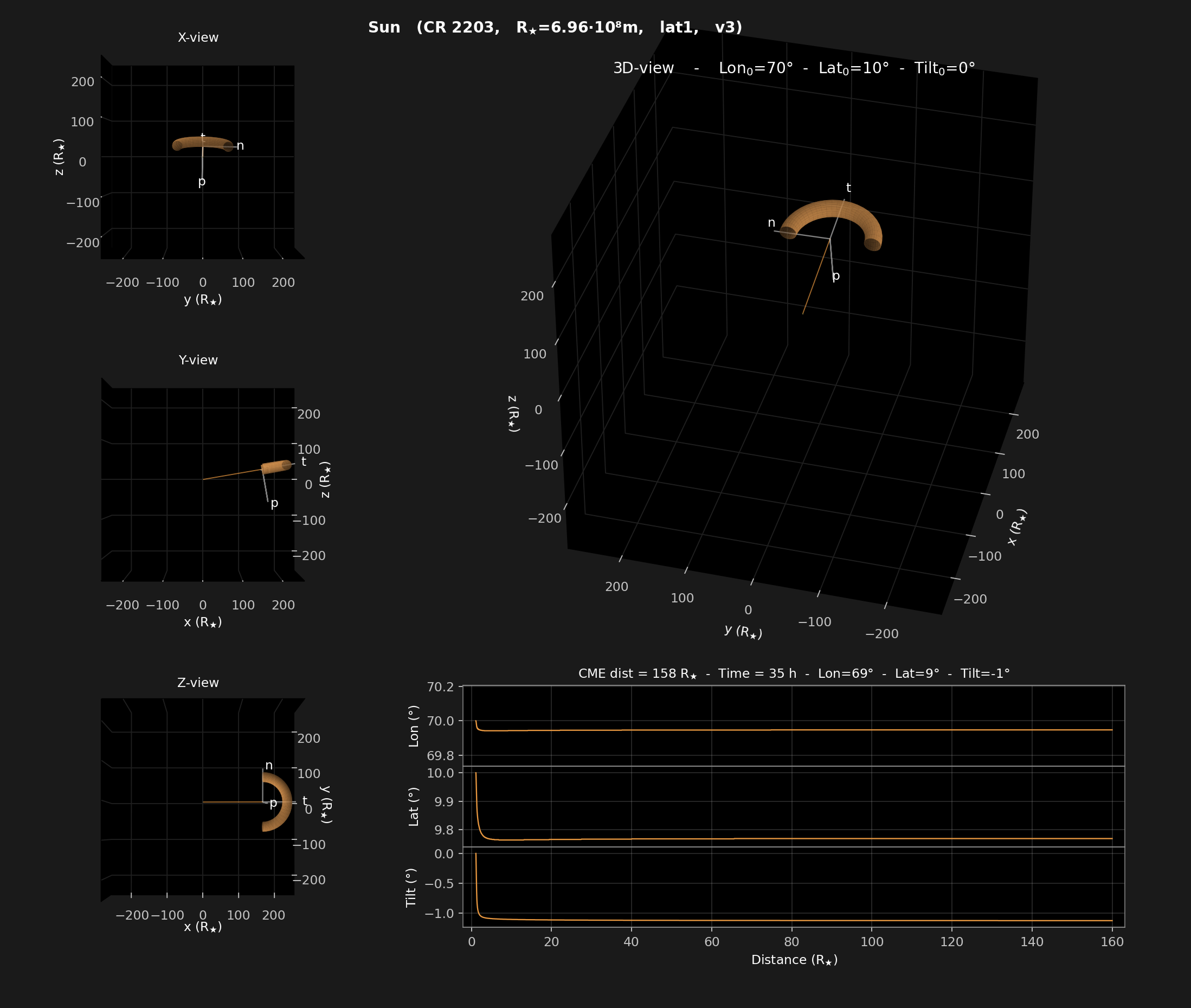}
\caption{Same as Figure~\ref{fig:3D-SUN-lat1-v1}, but for \textbf{Sun} model, \textbf{low-latitude}, \textbf{high-velocity}. The video version can be accessed via the links in the Supplementary Material section or \href{https://www.youtube.com/watch?v=qwArp5KrEr0}{this link}.}
\label{fig:3D-SUN-lat1-v3} \end{figure}

\begin{figure}
\centering
\includegraphics[width=\columnwidth]{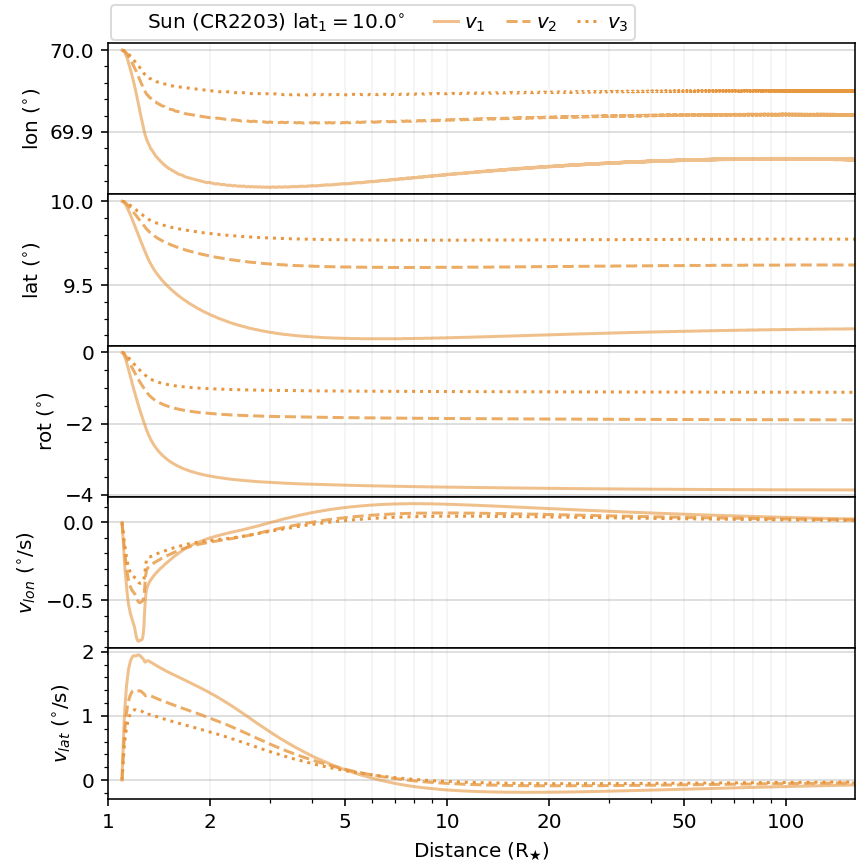}
\caption{ForeCAT output parameters of the CME trajectory as a function of radial distance (in $R_\star$, logarithmic scale), for the \textbf{Sun}, with \textbf{low-latitude} initial position. From top to bottom panels respectively: longitude variation, latitude variation, rotation, longitudinal velocity, and latitudinal velocity. In all panels, solid lines represent low-velocity  simulations; the dashed lines, mid-velocity simulations; and the dotted lines, the high-velocity simulations.}
\label{fig:SUN_l1} \end{figure}

\begin{figure}
\centering
\includegraphics[width=.99\columnwidth]{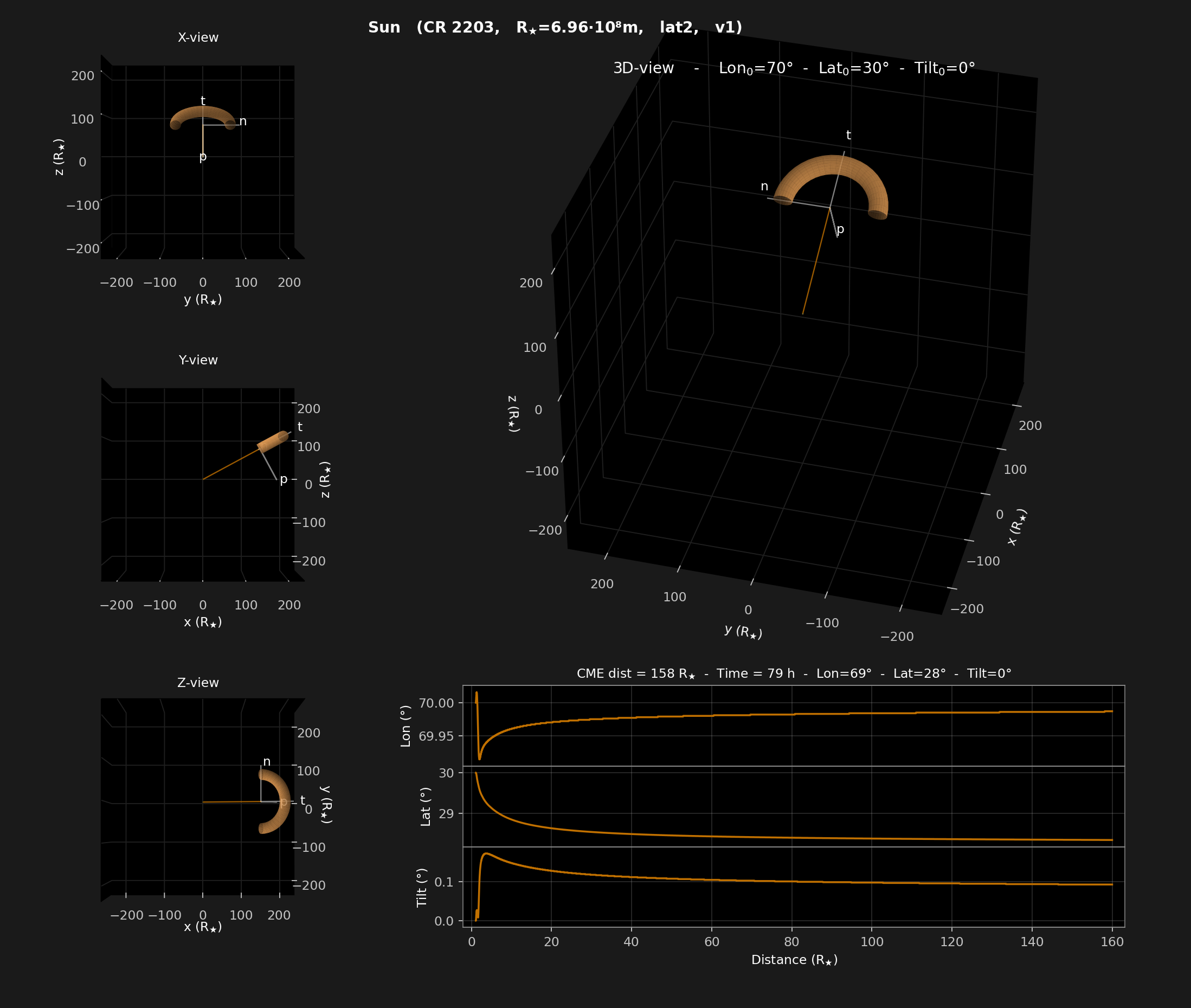}
\caption{Same as Figure~\ref{fig:3D-SUN-lat1-v1}, but for \textbf{Sun} model, \textbf{mid-latitude}, \textbf{low-velocity}. The video version can be accessed via the links in the Supplementary Material section or \href{https://www.youtube.com/watch?v=ebF3PHP9Tdc}{this link}.}
\label{fig:3D-SUN-lat2-v1} \end{figure}

\begin{figure}
\centering
\includegraphics[width=.99\columnwidth]{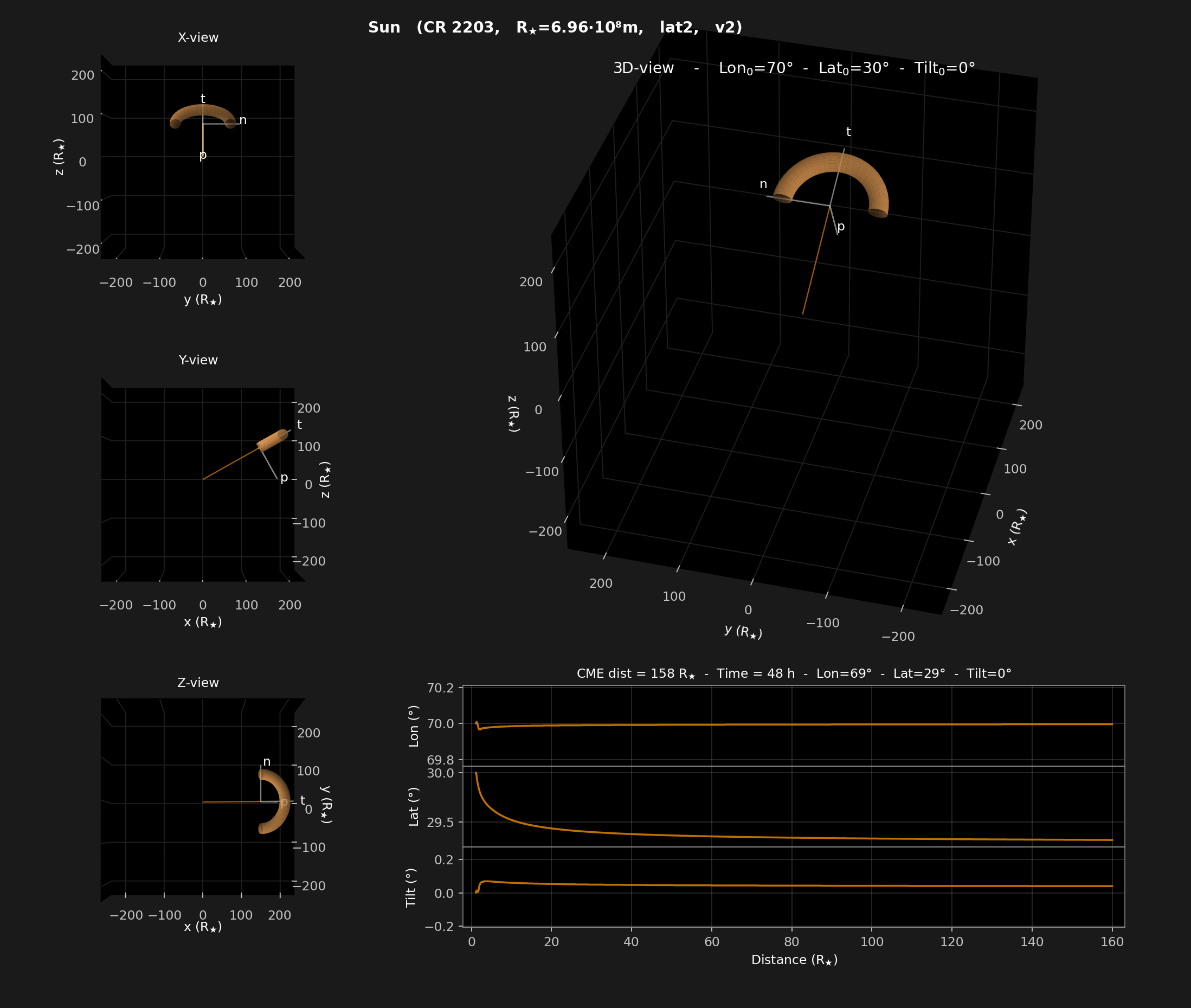}
\caption{Same as Figure~\ref{fig:3D-SUN-lat1-v1}, but for \textbf{Sun} model, \textbf{mid-latitude}, \textbf{mid-velocity}. The video version can be accessed via the links in the Supplementary Material section or \href{https://www.youtube.com/watch?v=qi8YYjKg8rk}{this link}.}
\label{fig:3D-SUN-lat2-v2} \end{figure}

\begin{figure}
\centering
\includegraphics[width=.99\columnwidth]{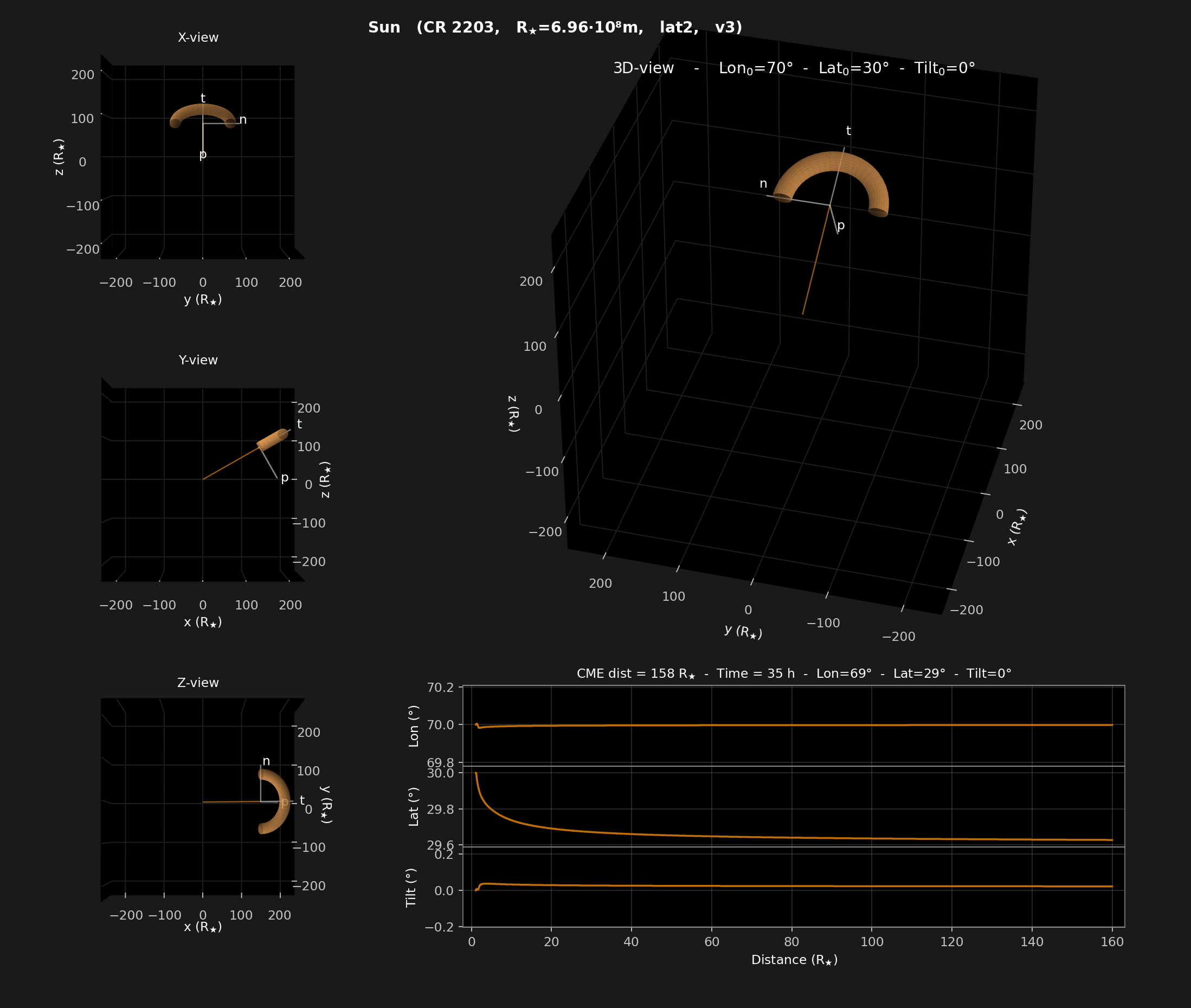}
\caption{Same as Figure~\ref{fig:3D-SUN-lat1-v1}, but for \textbf{Sun} model, \textbf{mid-latitude}, \textbf{high-velocity}. The video version can be accessed via the links in the Supplementary Material section or \href{https://www.youtube.com/watch?v=X2G7f18nhso}{this link}.}
\label{fig:3D-SUN-lat2-v3} \end{figure}

\begin{figure}
\centering
\includegraphics[width=\columnwidth]{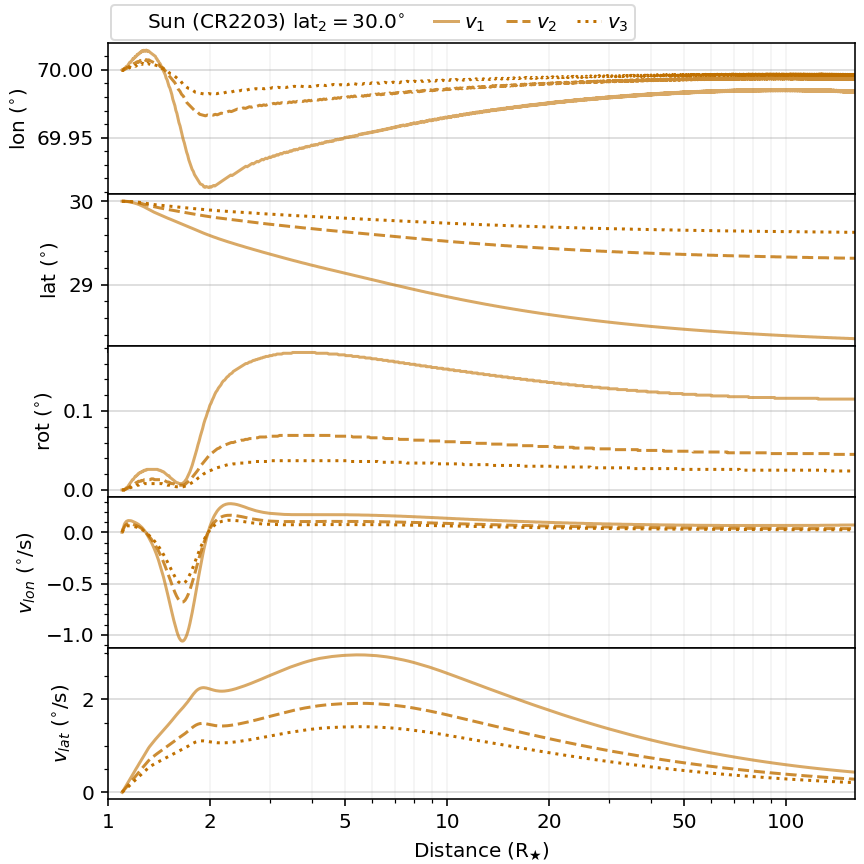}
\caption{Same as Figure~\ref{fig:SUN_l1}, but for \textbf{mid-latitude} simulations.}
\label{fig:SUN_l2} \end{figure}

\begin{figure}
\centering
\includegraphics[width=.99\columnwidth]{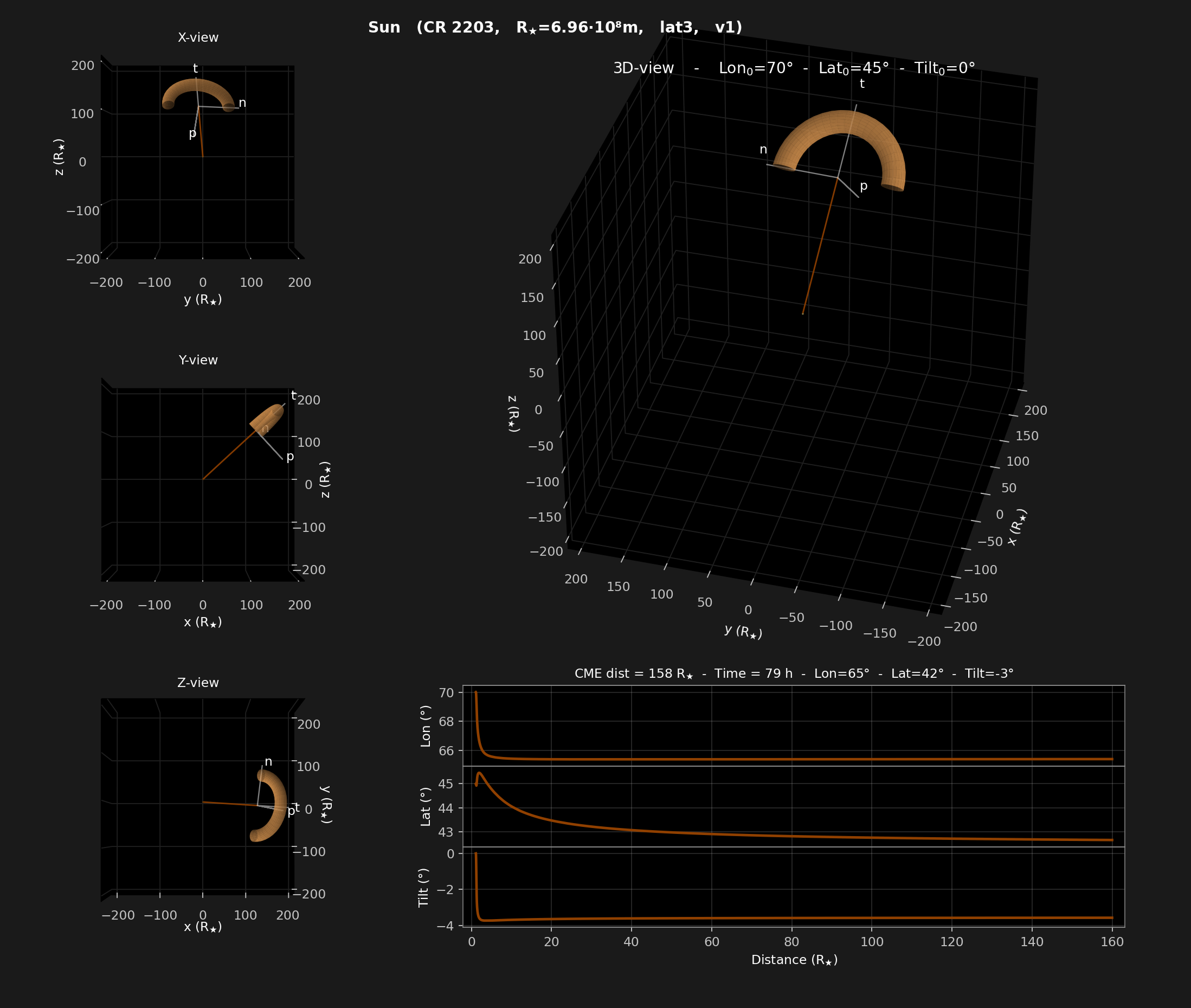}
\caption{Same as Figure~\ref{fig:3D-SUN-lat1-v1}, but for \textbf{Sun} model, \textbf{high-latitude}, \textbf{low-velocity}. The video version can be accessed via the links in the Supplementary Material section or \href{https://www.youtube.com/watch?v=JfP8UecuiDM}{this link}.}
\label{fig:3D-SUN-lat3-v1} \end{figure}

\begin{figure}
\centering
\includegraphics[width=.99\columnwidth]{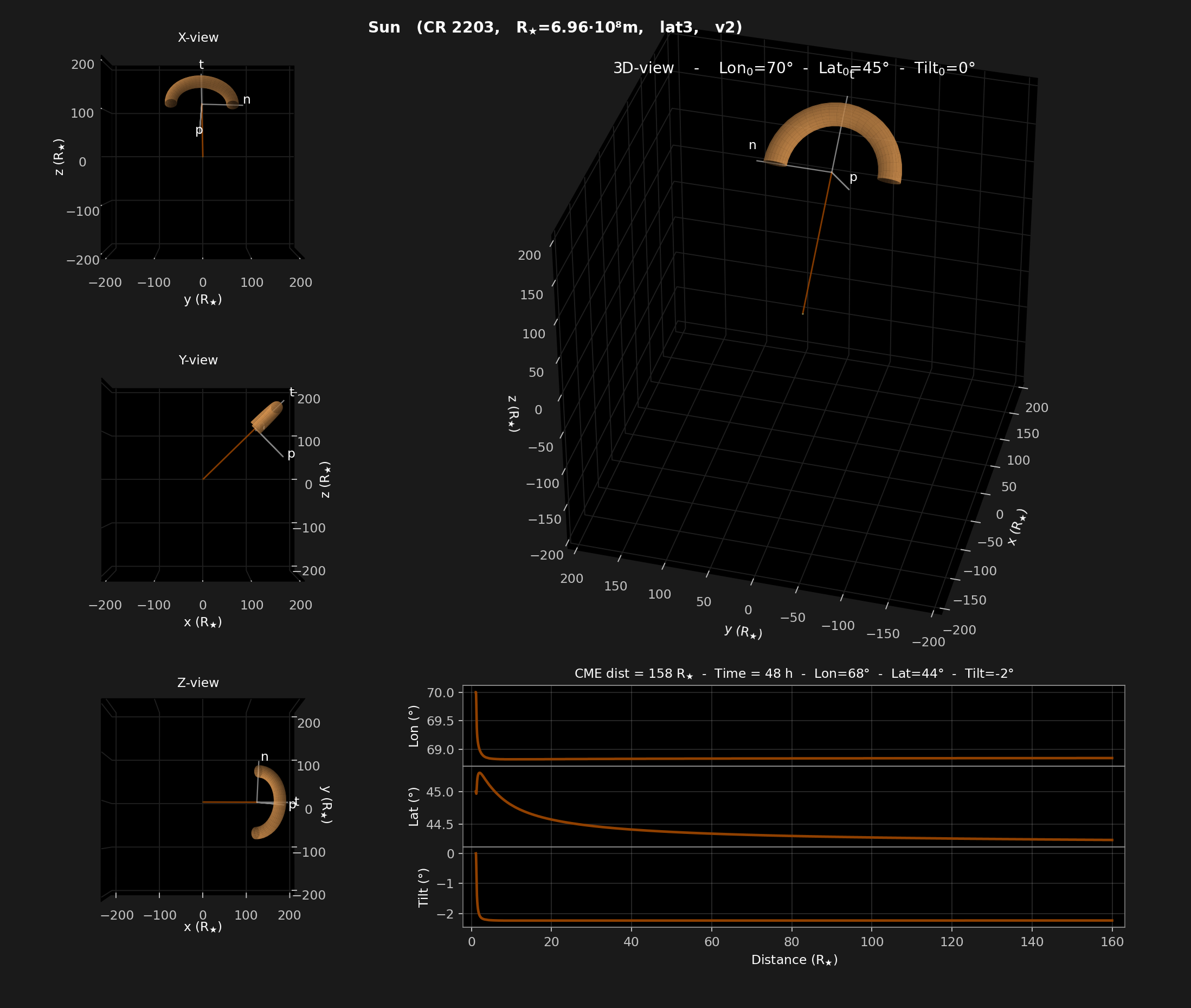}
\caption{Same as Figure~\ref{fig:3D-SUN-lat1-v1}, but for \textbf{Sun} model, \textbf{high-latitude}, \textbf{mid-velocity}. The video version can be accessed via the links in the Supplementary Material section or \href{https://www.youtube.com/watch?v=LJYkZ1u6Wk4}{this link}.}
\label{fig:3D-SUN-lat3-v2} \end{figure}

\begin{figure}
\centering
\includegraphics[width=.99\columnwidth]{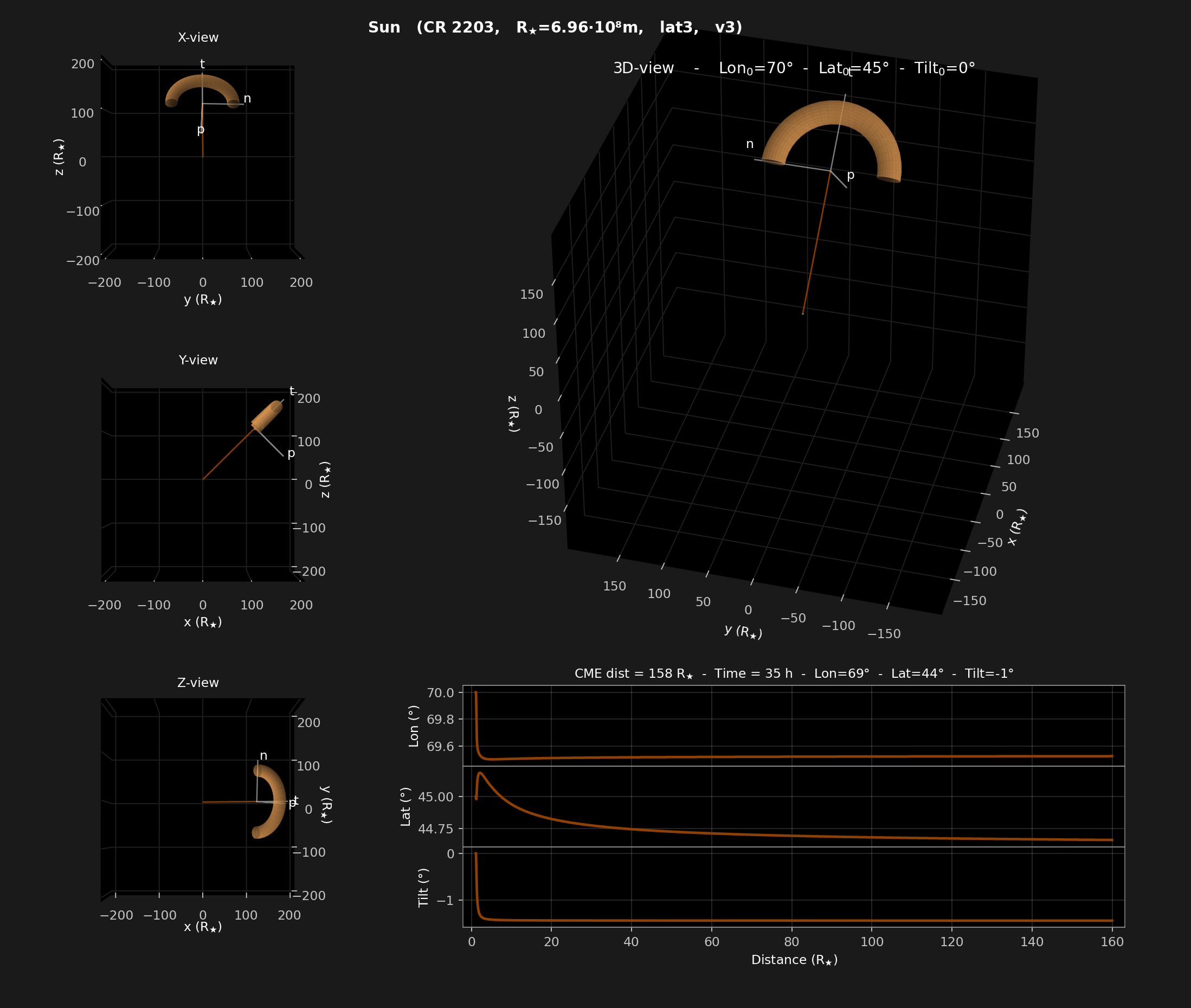}
\caption{Same as Figure~\ref{fig:3D-SUN-lat1-v1}, but for \textbf{Sun} model, \textbf{high-latitude}, \textbf{high-velocity}. The video version can be accessed via the links in the Supplementary Material section or \href{https://www.youtube.com/watch?v=__n9ylcaxKQ}{this link}.}
\label{fig:3D-SUN-lat3-v3} \end{figure}

\begin{figure}
\centering
\includegraphics[width=\columnwidth]{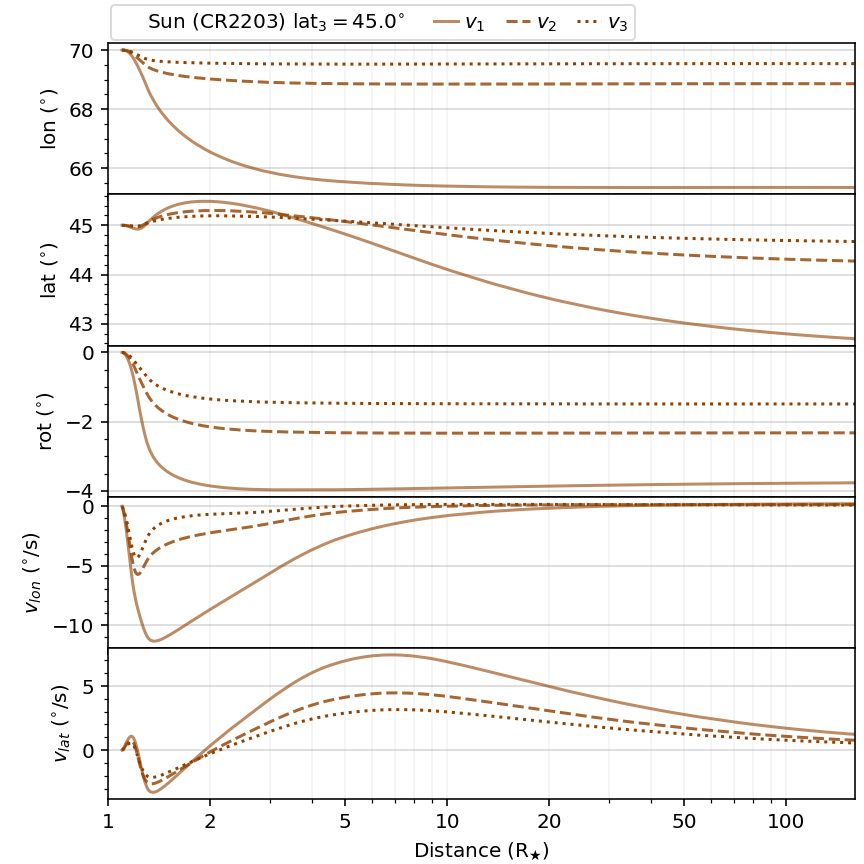}
\caption{Same as Figure~\ref{fig:SUN_l1}, but for \textbf{high-latitude} simulations.}
\label{fig:SUN_l3} \end{figure}

\begin{figure}
\centering
\includegraphics[width=.99\columnwidth]{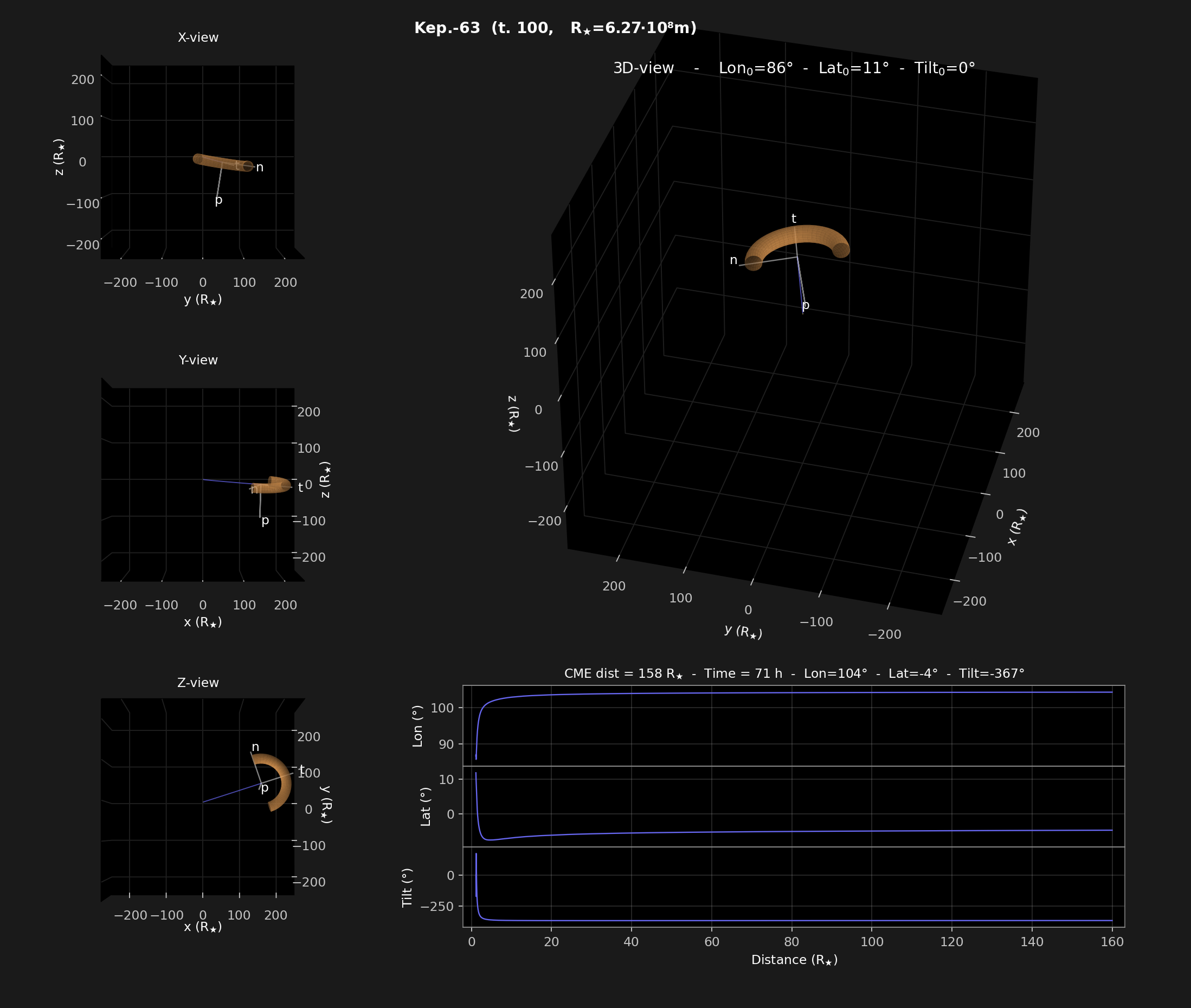}
\caption{Same as Figure~\ref{fig:3D-SUN-lat1-v1}, but for \textbf{Kepler-63} model, \textbf{high-latitude}, \textbf{low-velocity}. The video version can be accessed via the links in the Supplementary Material section or \href{https://www.youtube.com/watch?v=evcpylGc5Mo}{this link}.}
\label{fig:3D-K63-lat1-v1} \end{figure}

\begin{figure}
\centering
\includegraphics[width=.99\columnwidth]{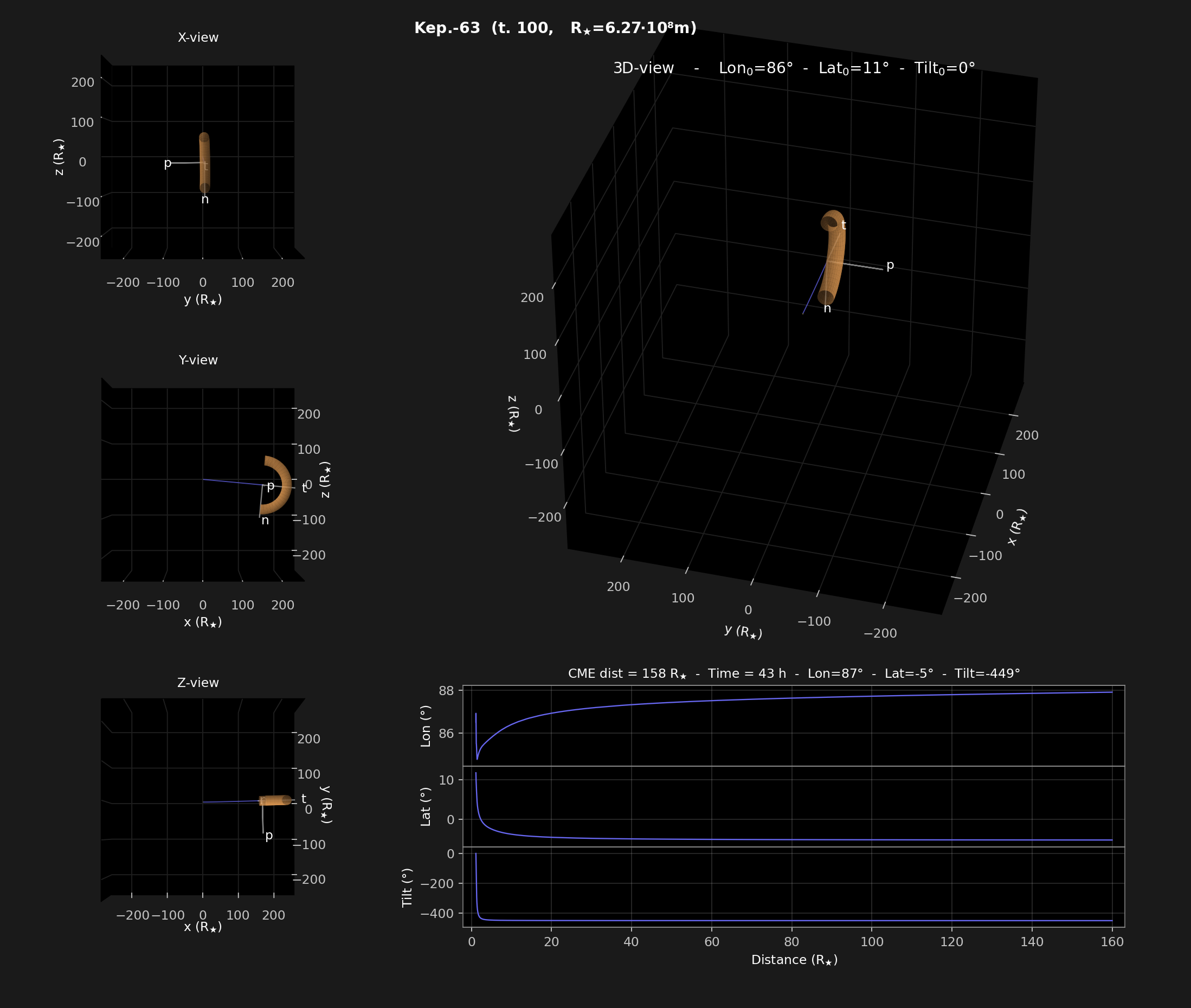}
\caption{Same as Figure~\ref{fig:3D-SUN-lat1-v1}, but for \textbf{Kepler-63} model, \textbf{high-latitude}, \textbf{mid-velocity}. The video version can be accessed via the links in the Supplementary Material section or \href{https://www.youtube.com/watch?v=EEr3O1SGvHI}{this link}.}
\label{fig:3D-K63-lat1-v2} \end{figure}

\begin{figure}
\centering
\includegraphics[width=.99\columnwidth]{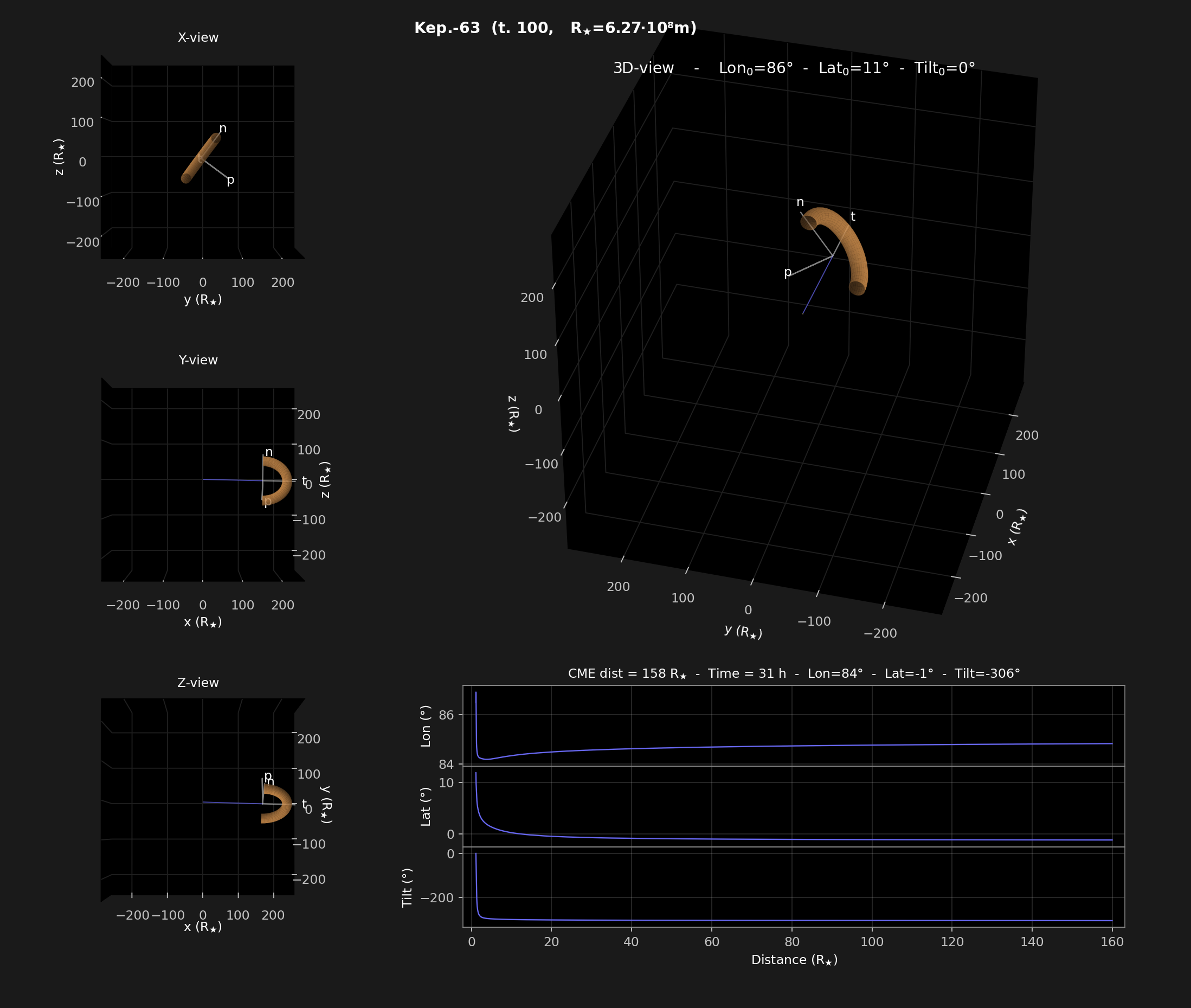}
\caption{Same as Figure~\ref{fig:3D-SUN-lat1-v1}, but for \textbf{Kepler-63} model, \textbf{high-latitude}, \textbf{high-velocity}. The video version can be accessed via the links in the Supplementary Material section or \href{https://www.youtube.com/watch?v=opF6yNo8mvc}{this link}.}
\label{fig:3D-K63-lat1-v3} \end{figure}

\begin{figure}
\centering
\includegraphics[width=\columnwidth]{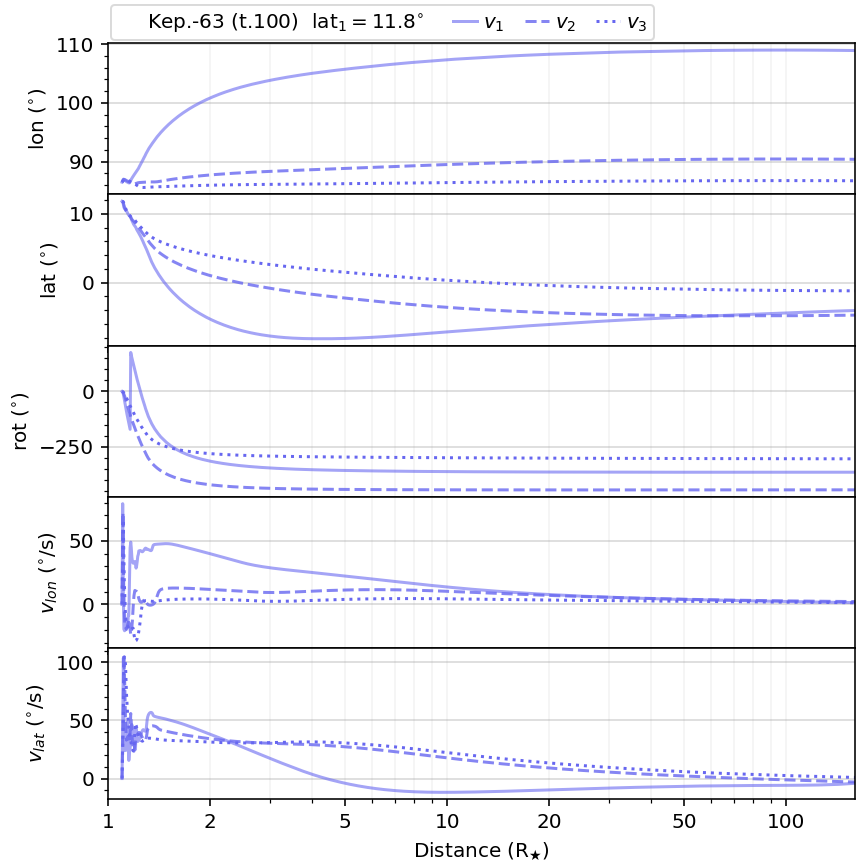}
\caption{Same as Figure~\ref{fig:SUN_l1}, but for \textbf{Kepler-63}, \textbf{low-latitude} simulations.}
\label{fig:K63_l1} \end{figure}

\begin{figure}
\centering
\includegraphics[width=.99\columnwidth]{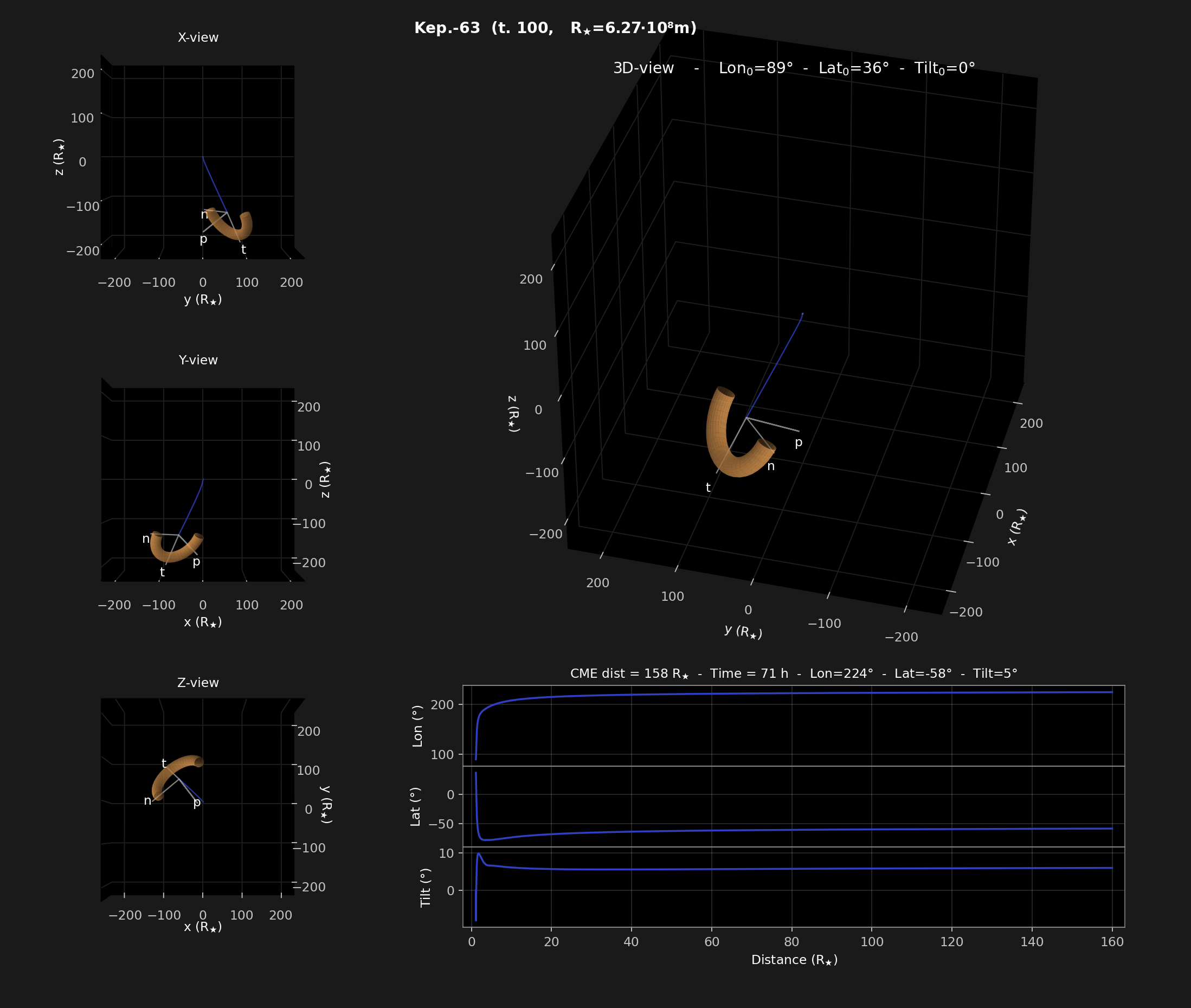}
\caption{Same as Figure~\ref{fig:3D-SUN-lat1-v1}, but for \textbf{Kepler-63} model, \textbf{mid-latitude}, \textbf{low-velocity}. The video version can be accessed via the links in the Supplementary Material section or \href{https://www.youtube.com/watch?v=hmpZk5H34Tk}{this link}.}
\label{fig:3D-K63-lat2-v1} \end{figure}

\begin{figure}
\centering
\includegraphics[width=.99\columnwidth]{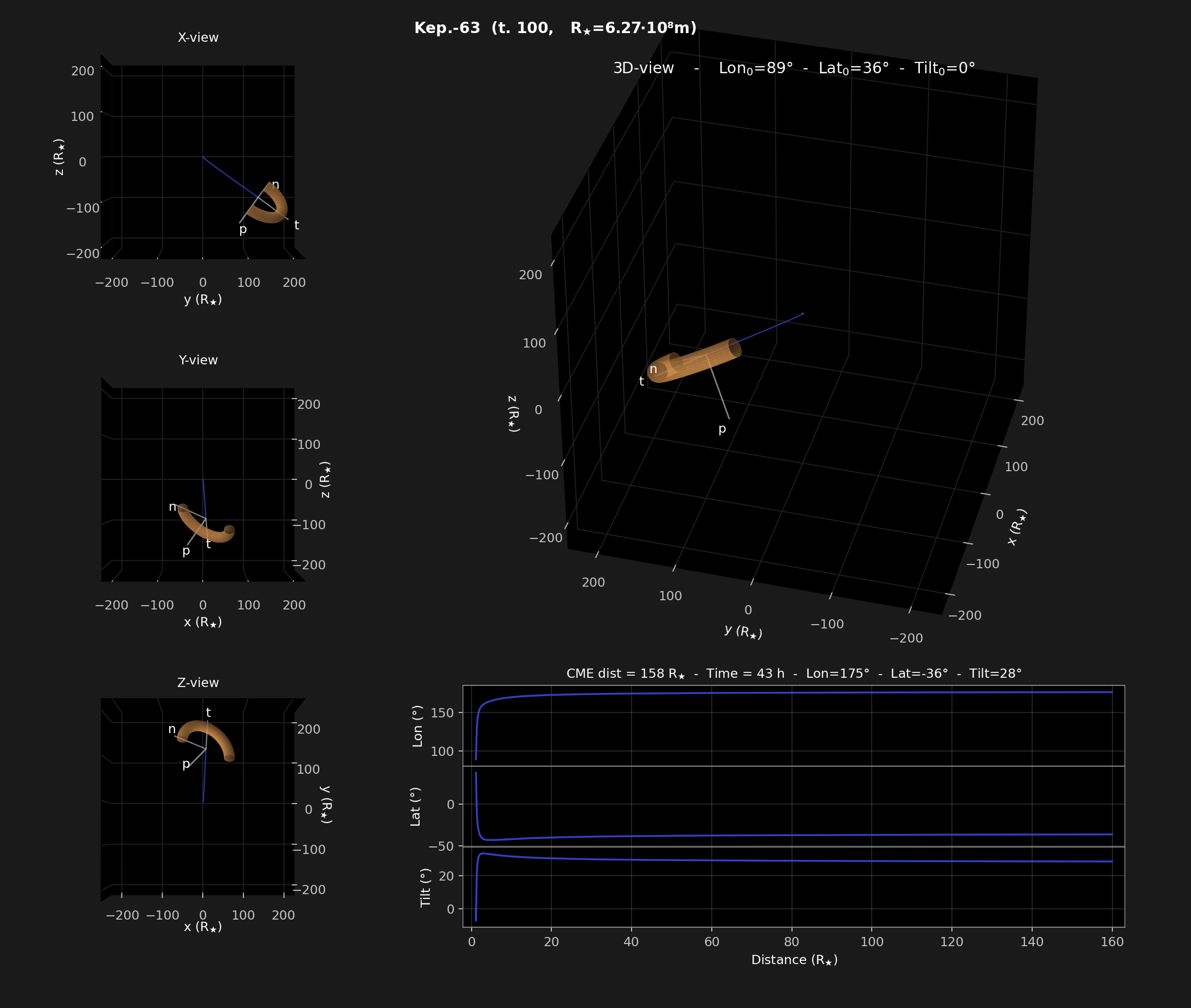}
\caption{Same as Figure~\ref{fig:3D-SUN-lat1-v1}, but for \textbf{Kepler-63} model, \textbf{mid-latitude}, \textbf{mid-velocity}. The video version can be accessed via the links in the Supplementary Material section or \href{https://www.youtube.com/watch?v=mc-uTvt1Cig}{this link}.}
\label{fig:3D-K63-lat2-v2} \end{figure}

\begin{figure}
\centering
\includegraphics[width=.99\columnwidth]{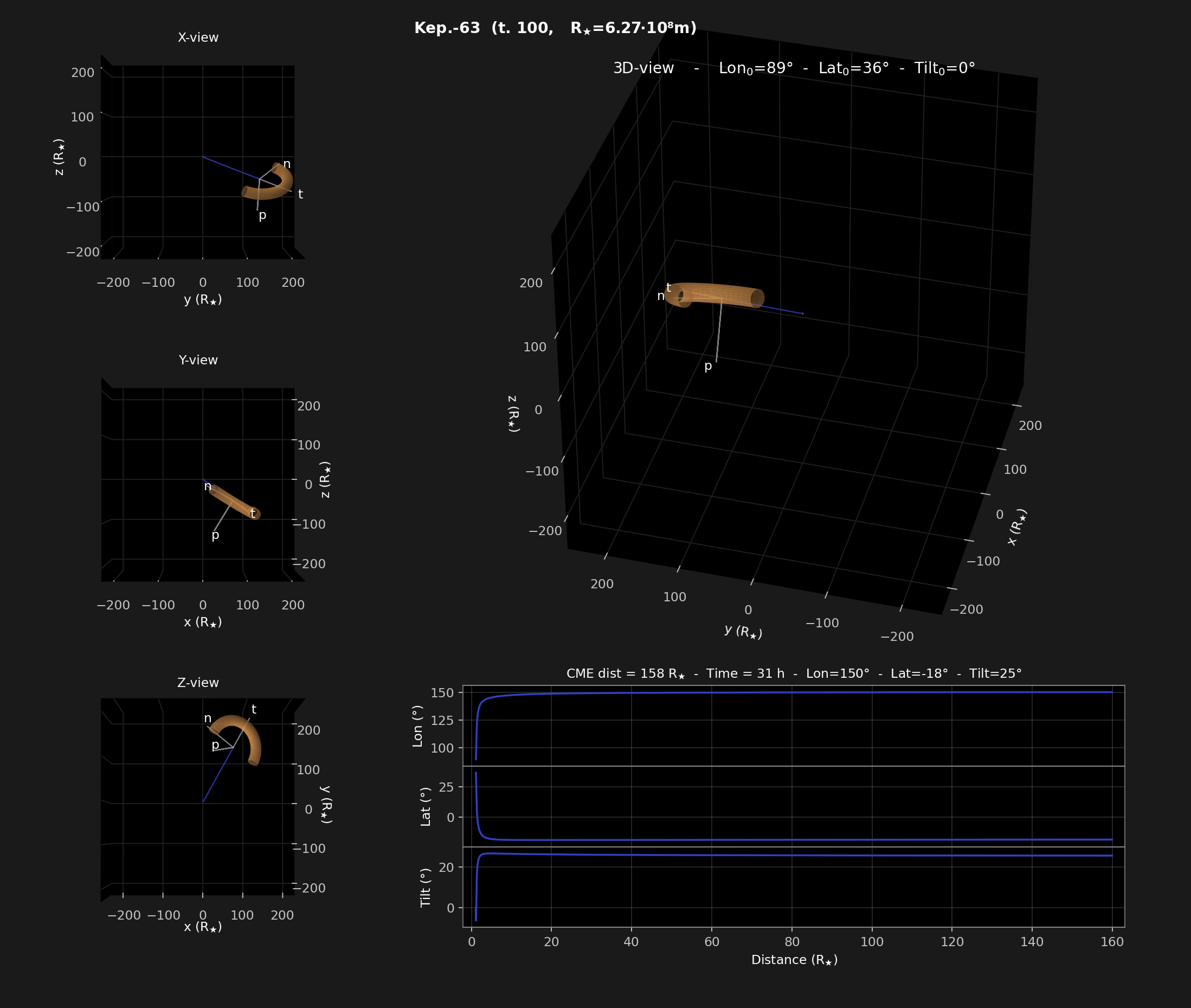}
\caption{Same as Figure~\ref{fig:3D-SUN-lat1-v1}, but for \textbf{Kepler-63} model, \textbf{mid-latitude}, \textbf{high-velocity}. The video version can be accessed via the links in the Supplementary Material section or \href{https://www.youtube.com/watch?v=Khvlhm6SwSo}{this link}.}
\label{fig:3D-K63-lat2-v3} \end{figure}

\begin{figure}
\centering
\includegraphics[width=\columnwidth]{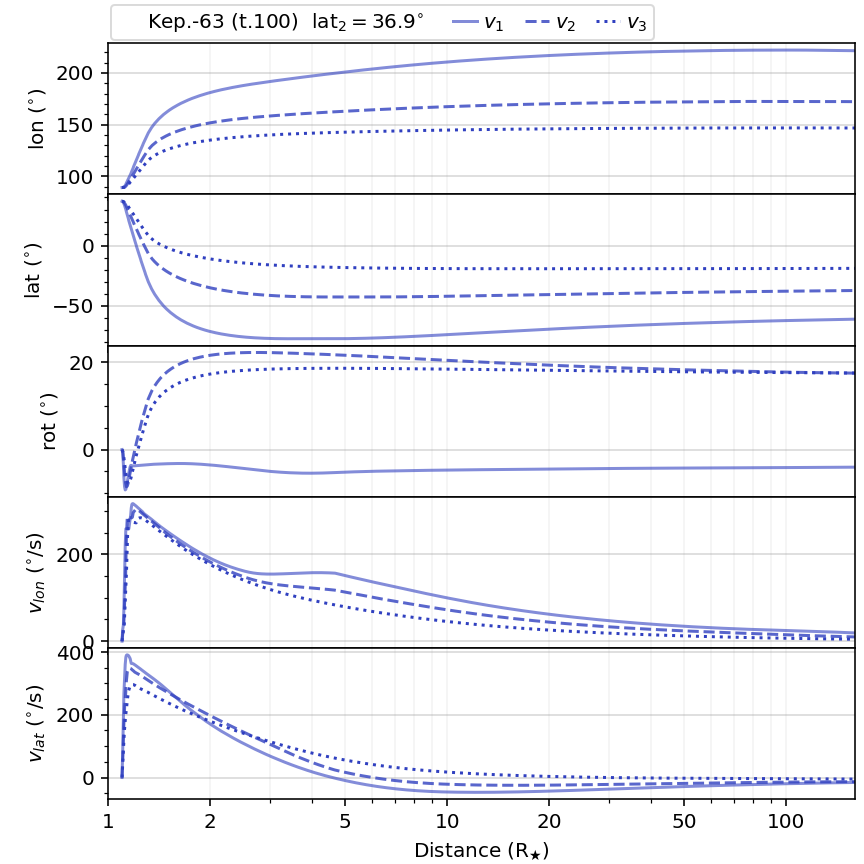}
\caption{Same as Figure~\ref{fig:SUN_l1}, but for \textbf{Kepler-63}, \textbf{mid-latitude} simulations.}
\label{fig:K63_l2} \end{figure}

\begin{figure}
\centering
\includegraphics[width=.99\columnwidth]{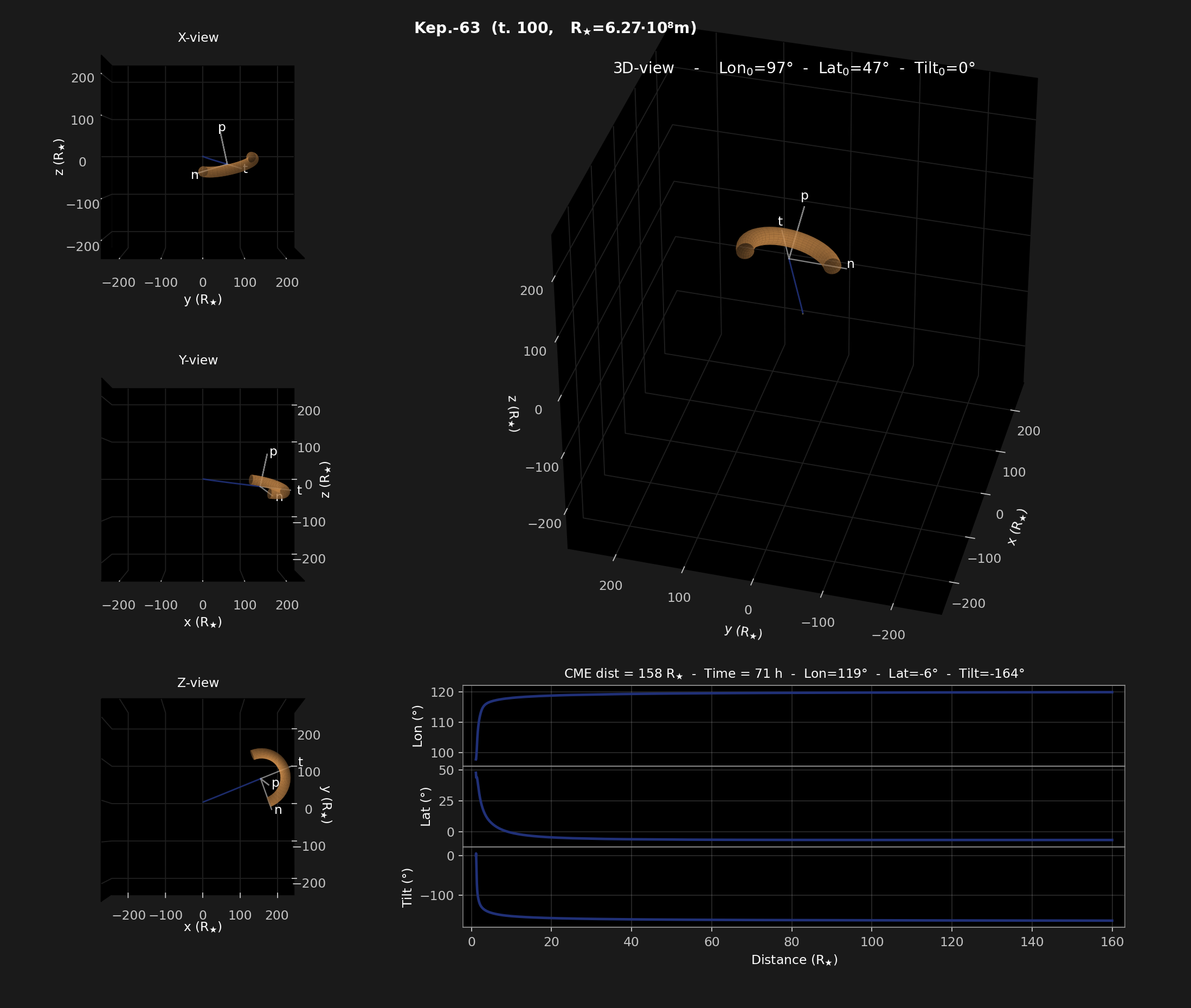}
\caption{Same as Figure~\ref{fig:3D-SUN-lat1-v1}, but for \textbf{Kepler-63} model, \textbf{high-latitude}, \textbf{low-velocity}. The video version can be accessed via the links in the Supplementary Material section or \href{https://www.youtube.com/watch?v=PdKq1jsblG4}{this link}.}
\label{fig:3D-K63-lat3-v1} \end{figure}

\begin{figure}
\centering
\includegraphics[width=.99\columnwidth]{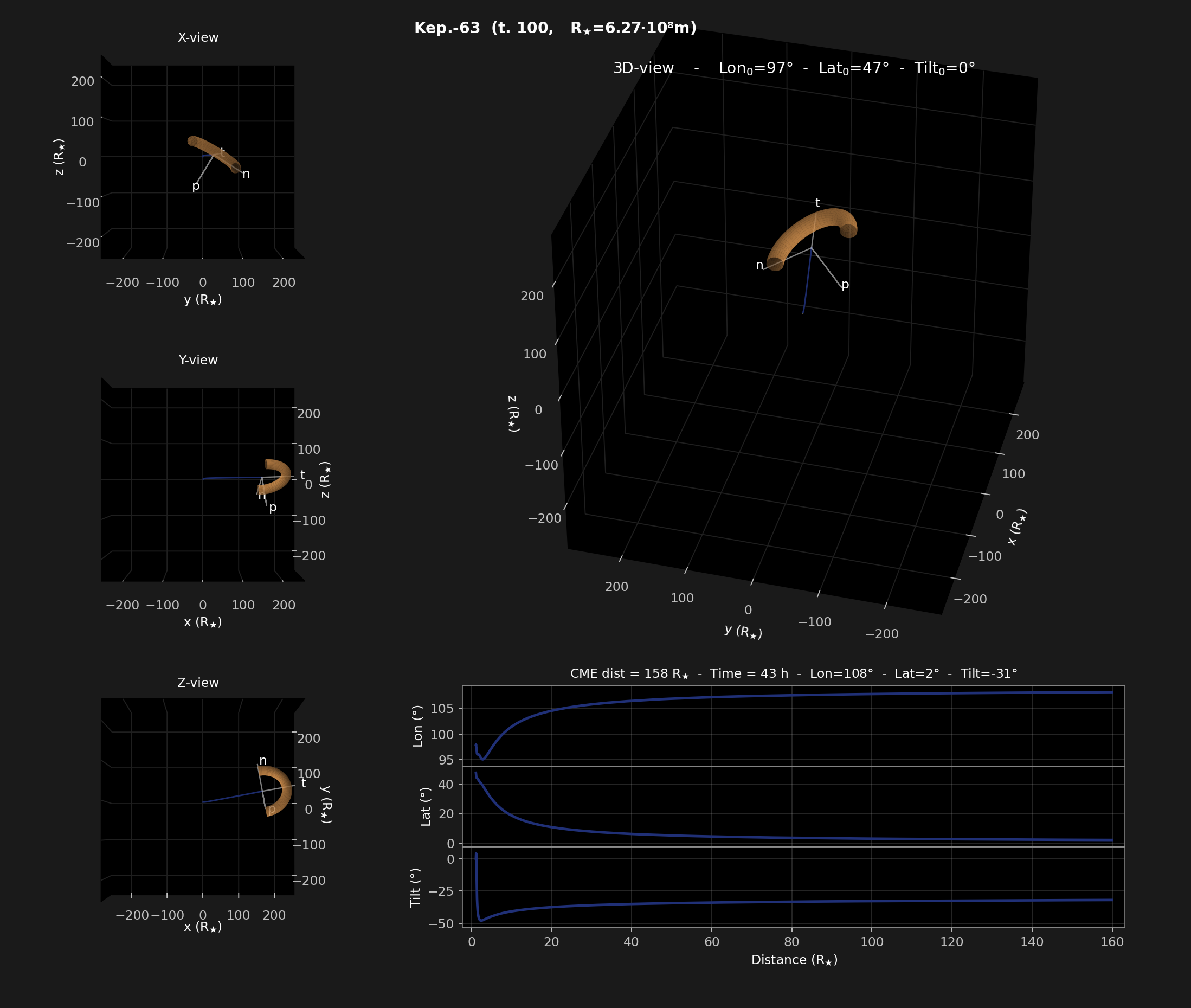}
\caption{Same as Figure~\ref{fig:3D-SUN-lat1-v1}, but for \textbf{Kepler-63} model, \textbf{high-latitude}, \textbf{mid-velocity}. The video version can be accessed via the links in the Supplementary Material section or \href{https://www.youtube.com/watch?v=KDIbz_42BVw}{this link}.}
\label{fig:3D-K63-lat3-v2} \end{figure}

\begin{figure}
\centering
\includegraphics[width=.99\columnwidth]{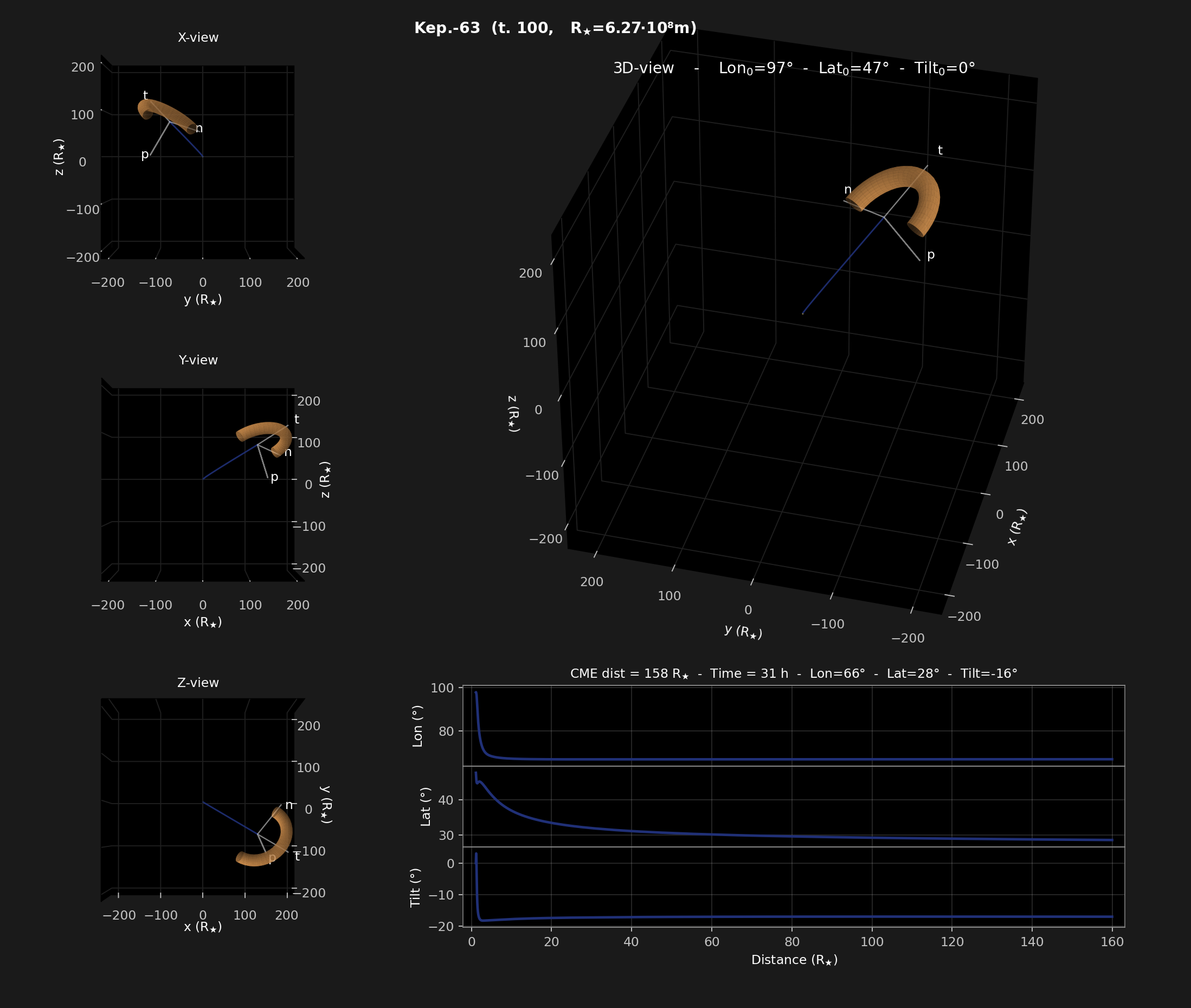}
\caption{Same as Figure~\ref{fig:3D-SUN-lat1-v1}, but for \textbf{Kepler-63} model, \textbf{high-latitude}, \textbf{high-velocity}. The video version can be accessed via the links in the Supplementary Material section or \href{https://www.youtube.com/watch?v=5d9efpzblxc}{this link}.}
\label{fig:3D-K63-lat3-v3} \end{figure}

\begin{figure}
\centering
\includegraphics[width=\columnwidth]{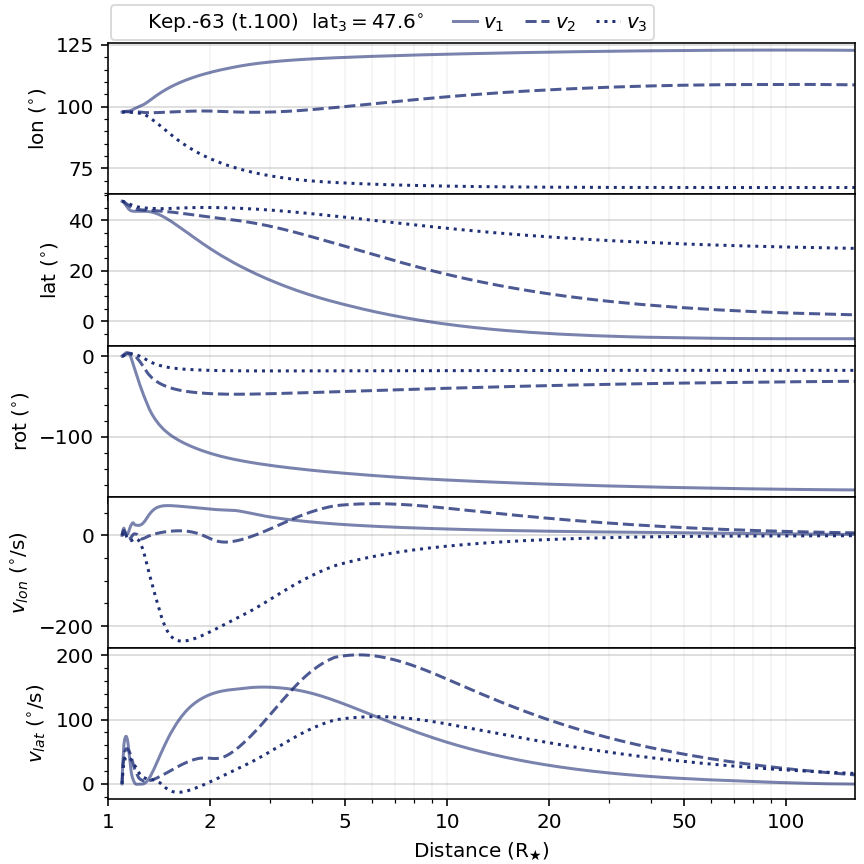}
\caption{Same as Figure~\ref{fig:SUN_l1}, but for \textbf{Kepler-63}, \textbf{high-latitude} simulations.}
\label{fig:K63_l3} \end{figure}

\begin{figure}
\centering
\includegraphics[width=.99\columnwidth]{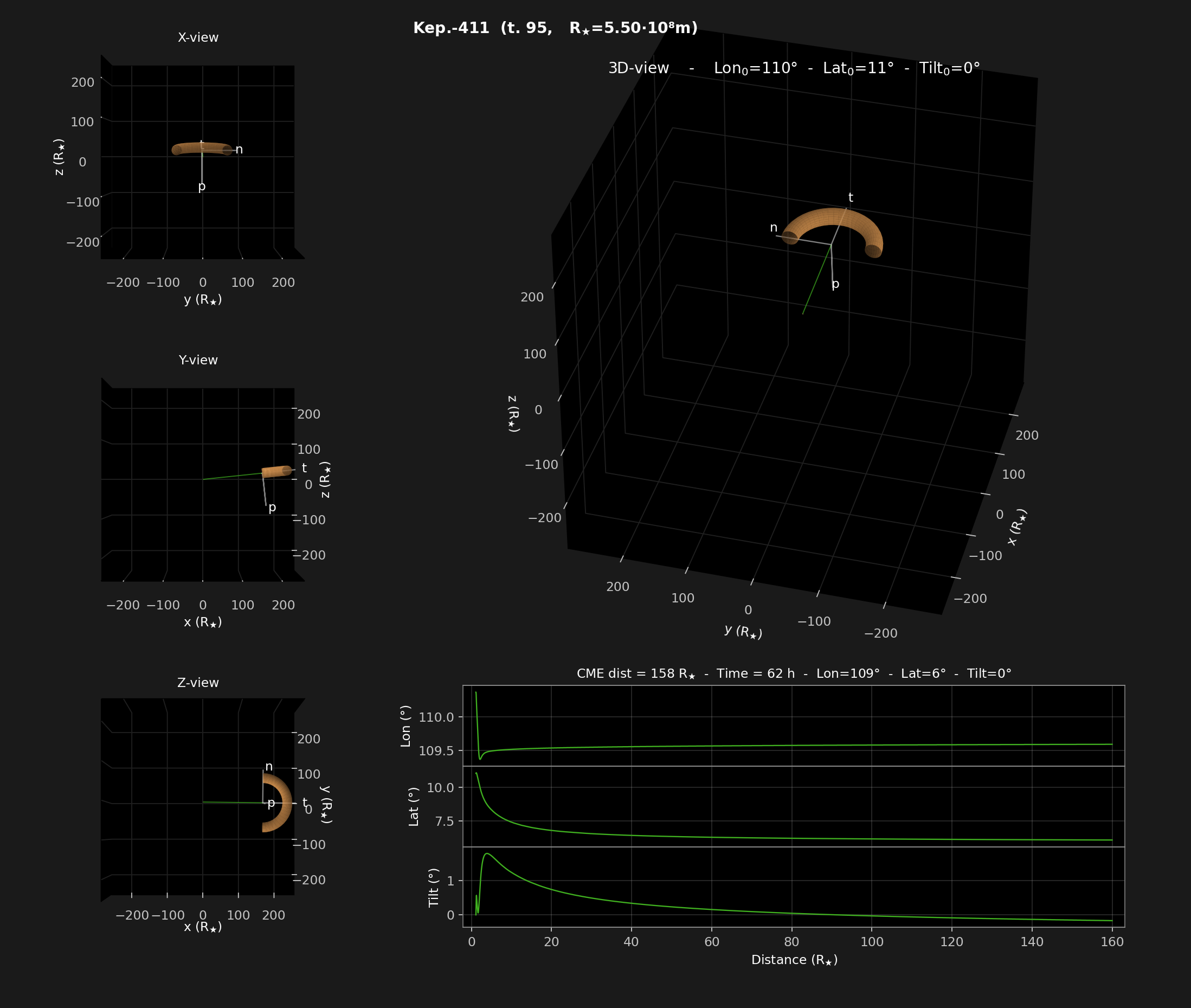}
\caption{Same as Figure~\ref{fig:3D-SUN-lat1-v1}, but for \textbf{Kepler-411} model, \textbf{low-latitude}, \textbf{low-velocity}. The video version can be accessed via the links in the Supplementary Material section or \href{https://www.youtube.com/watch?v=bQ0i7XbKEv8}{this link}.}
\label{fig:3D-K411-lat1-v1} \end{figure}

\begin{figure}
\centering
\includegraphics[width=.99\columnwidth]{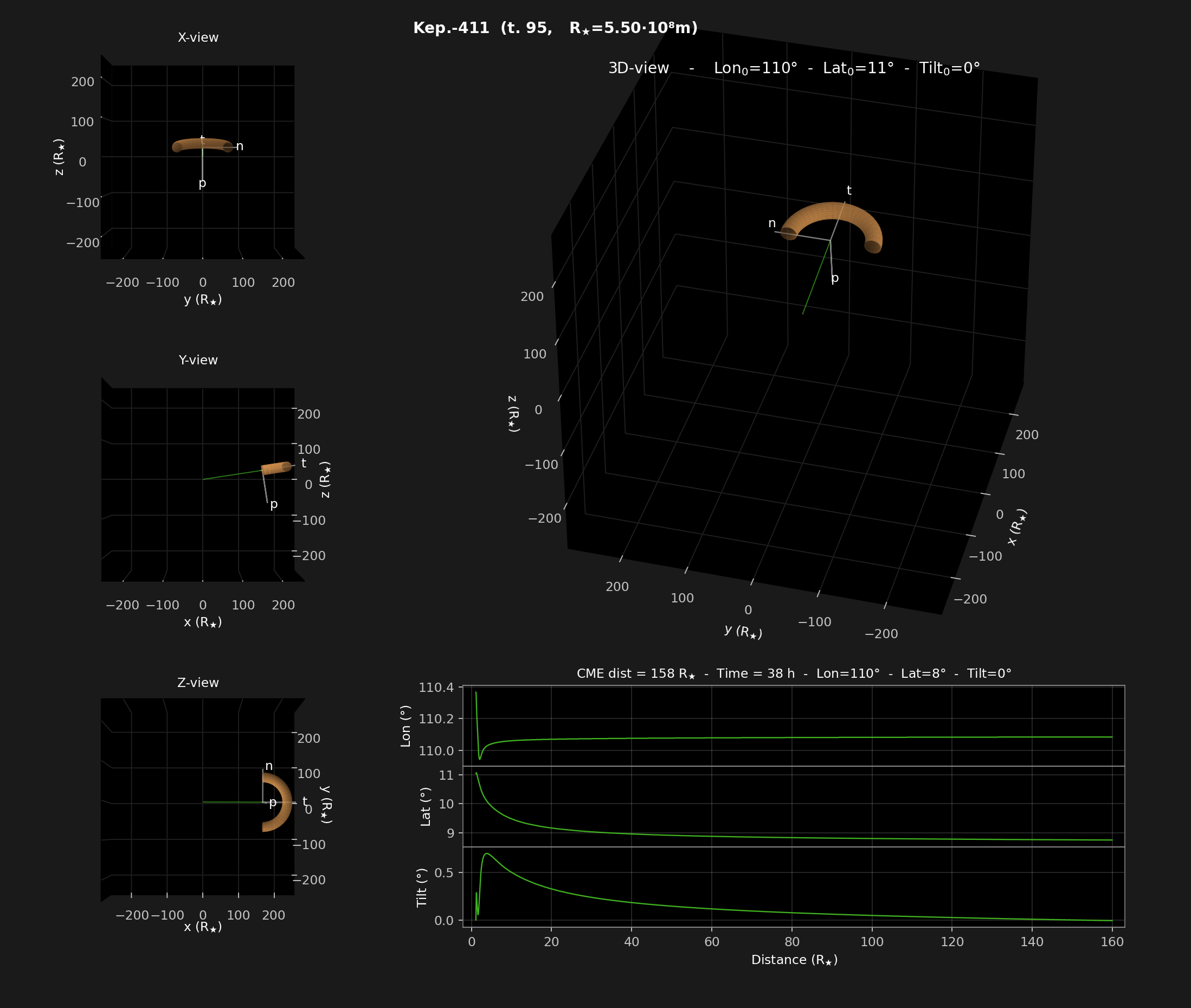}
\caption{Same as Figure~\ref{fig:3D-SUN-lat1-v1}, but for \textbf{Kepler-411} model, \textbf{low-latitude}, \textbf{mid-velocity}. The video version can be accessed via the links in the Supplementary Material section or \href{https://www.youtube.com/watch?v=Gafsgz2Lf9I}{this link}.}
\label{fig:3D-K411-lat1-v2} \end{figure}

\begin{figure}
\centering
\includegraphics[width=.99\columnwidth]{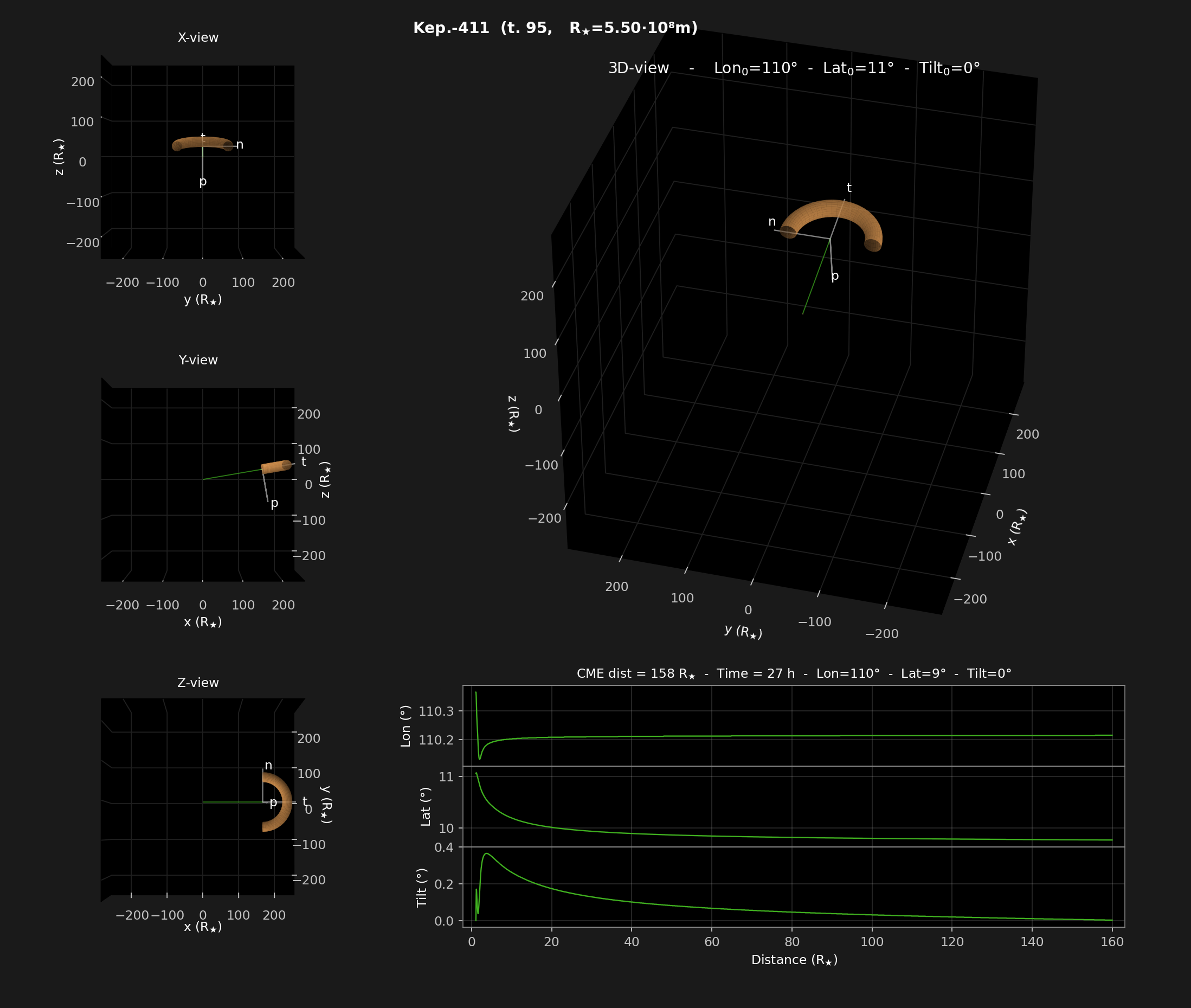}
\caption{Same as Figure~\ref{fig:3D-SUN-lat1-v1}, but for \textbf{Kepler-411} model, \textbf{low-latitude}, \textbf{high-velocity}. The video version can be accessed via the links in the Supplementary Material section or \href{https://www.youtube.com/watch?v=BjyyfWDiVxg}{this link}.}
\label{fig:3D-K411-lat1-v3} \end{figure}

\begin{figure}
\centering
\includegraphics[width=\columnwidth]{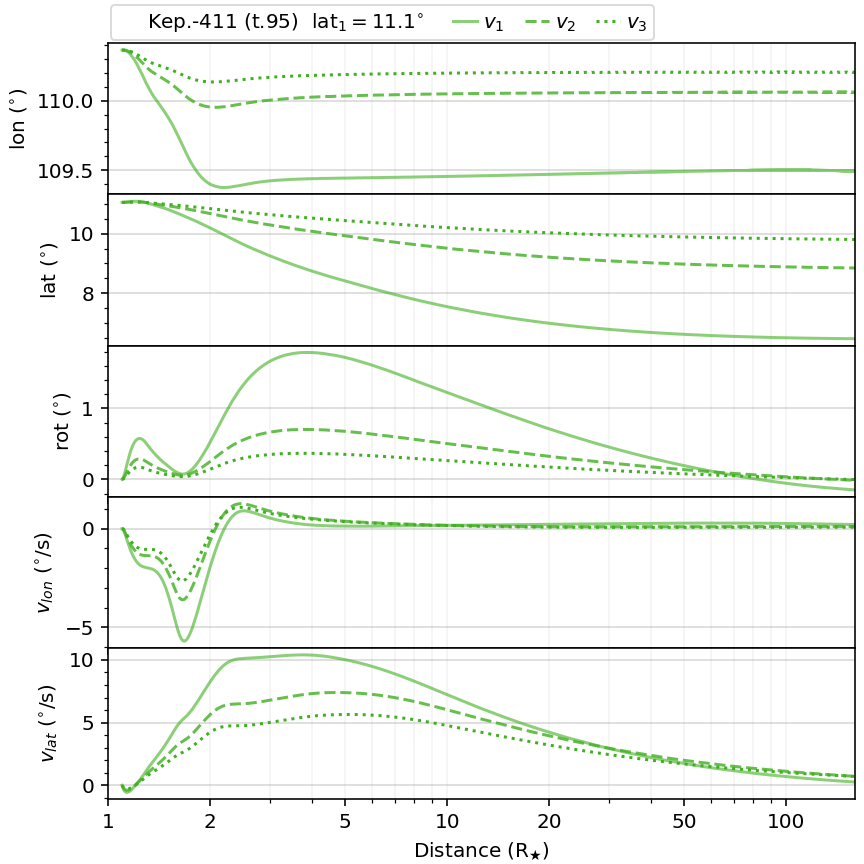}
\caption{Same as Figure~\ref{fig:SUN_l1}, but for \textbf{Kepler-411}, \textbf{low-latitude} simulations.}
\label{fig:K411_l1} \end{figure}

\begin{figure}
\centering
\includegraphics[width=.99\columnwidth]{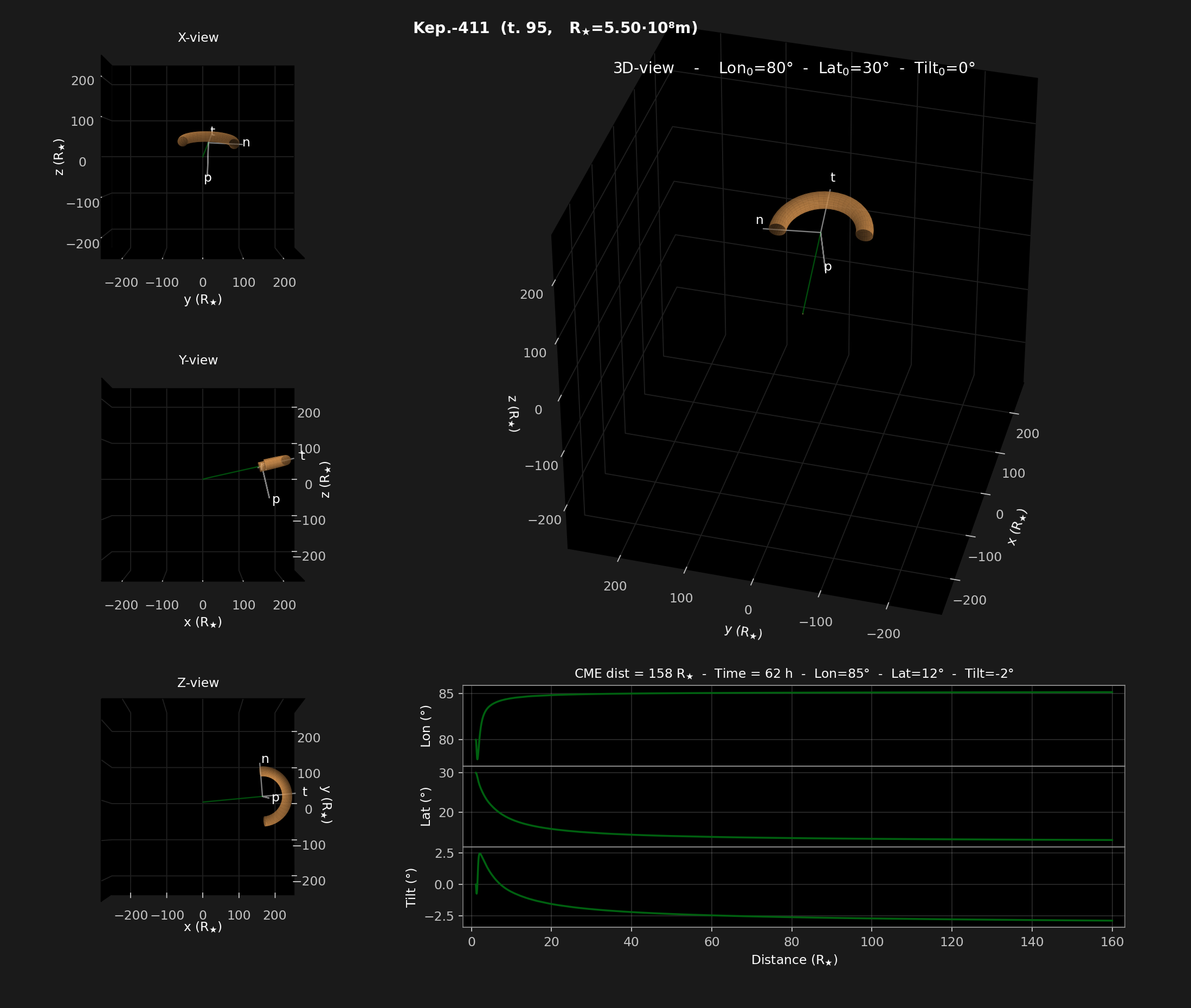}
\caption{Same as Figure~\ref{fig:3D-SUN-lat1-v1}, but for \textbf{Kepler-411} model, \textbf{mid-latitude}, \textbf{low-velocity}. The video version can be accessed via the links in the Supplementary Material section or \href{https://www.youtube.com/watch?v=EFAPnLOrHbU}{this link}.}
\label{fig:3D-K411-lat2-v1} \end{figure}

\begin{figure}
\centering
\includegraphics[width=.99\columnwidth]{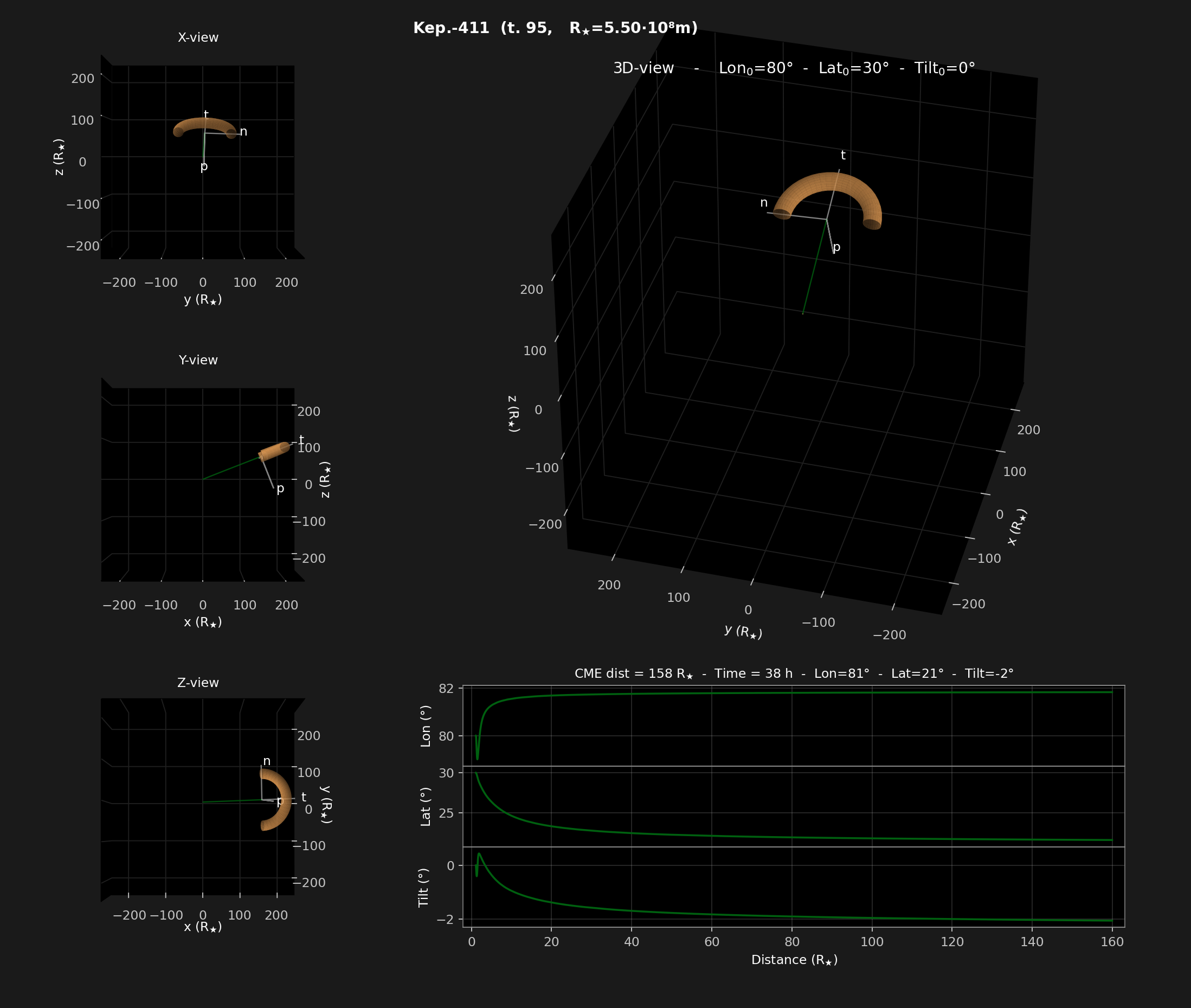}
\caption{Same as Figure~\ref{fig:3D-SUN-lat1-v1}, but for \textbf{Kepler-411} model, \textbf{mid-latitude}, \textbf{mid-velocity}. The video version can be accessed via the links in the Supplementary Material section or \href{https://www.youtube.com/watch?v=LUuOC5OSJN8}{this link}.}
\label{fig:3D-K411-lat2-v2} \end{figure}

\begin{figure}
\centering
\includegraphics[width=.99\columnwidth]{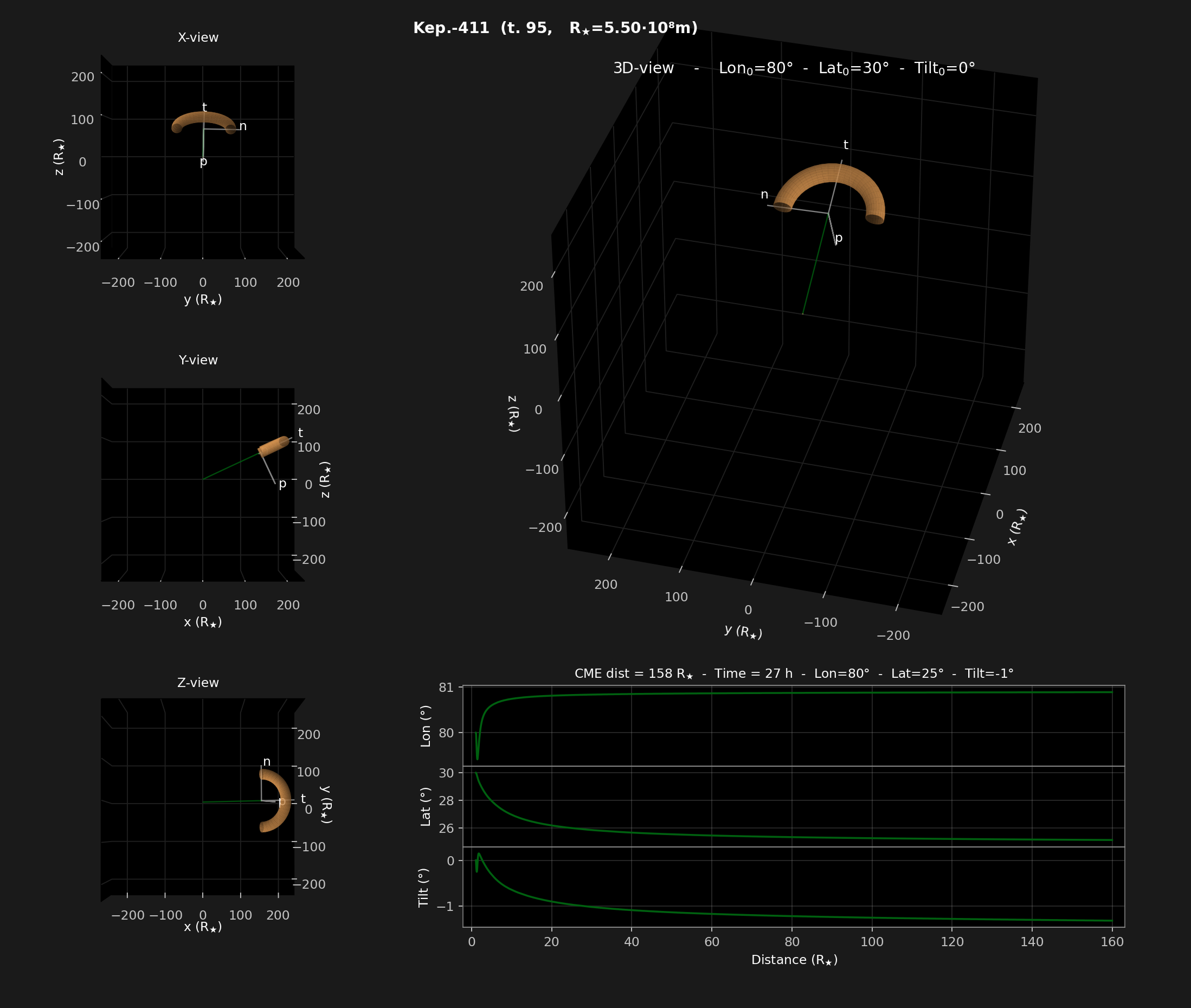}
\caption{Same as Figure~\ref{fig:3D-SUN-lat1-v1}, but for \textbf{Kepler-411} model, \textbf{mid-latitude}, \textbf{high-velocity}. The video version can be accessed via the links in the Supplementary Material section or \href{https://www.youtube.com/watch?v=ArJcb_yQIAQ}{this link}.}
\label{fig:3D-K411-lat2-v3} \end{figure}

\begin{figure}
\centering
\includegraphics[width=\columnwidth]{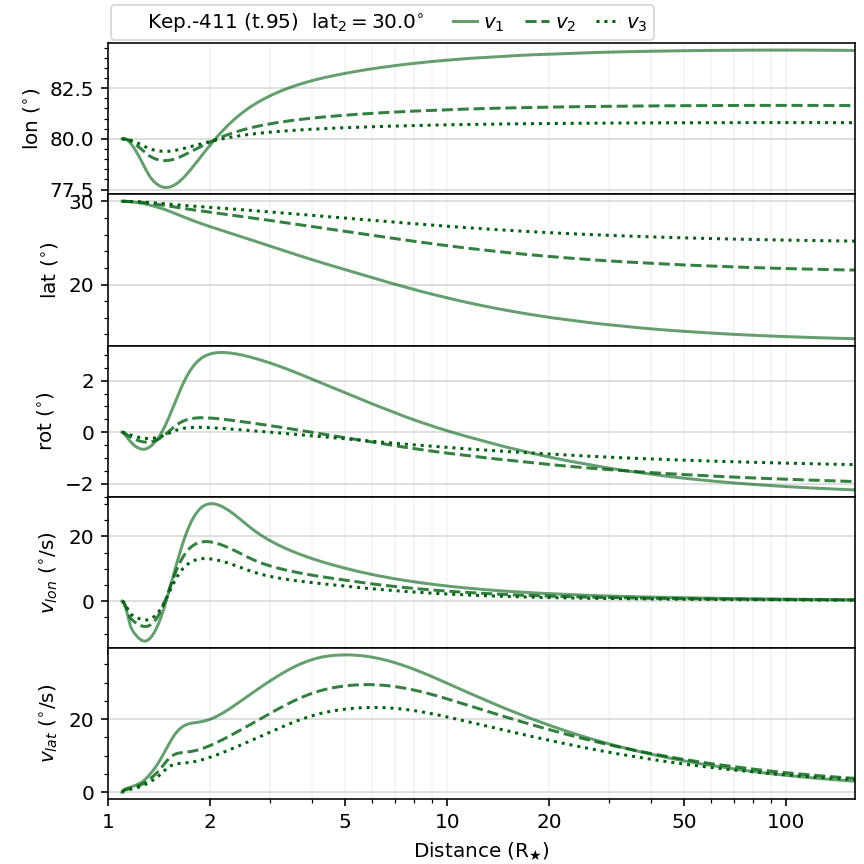}
\caption{Same as Figure~\ref{fig:SUN_l1}, but for \textbf{Kepler-411}, \textbf{mid-latitude} simulations.}
\label{fig:K411_l2} \end{figure}

\begin{figure}
\centering
\includegraphics[width=.99\columnwidth]{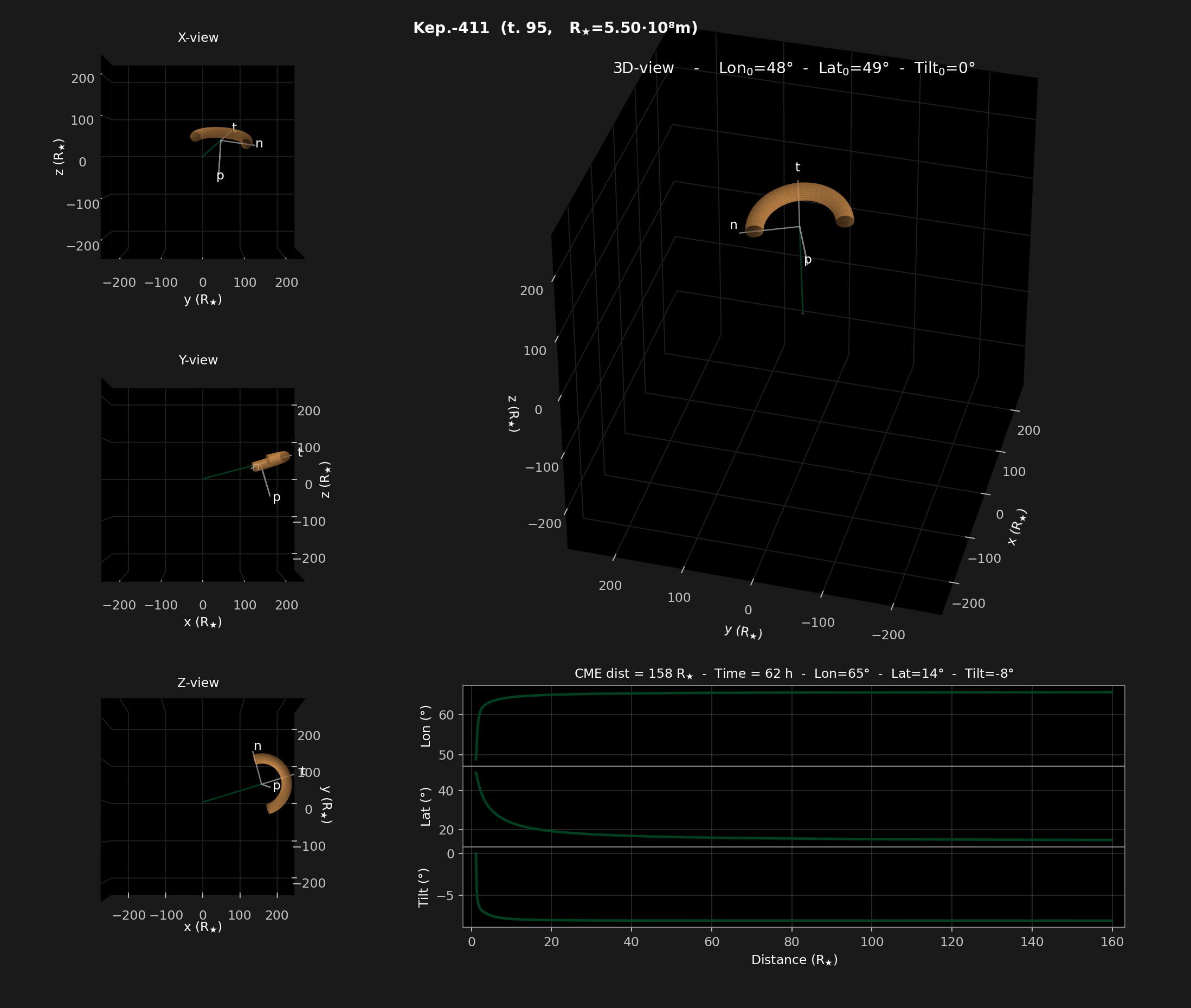}
\caption{Same as Figure~\ref{fig:3D-SUN-lat1-v1}, but for \textbf{Kepler-411} model, \textbf{high-latitude}, \textbf{low-velocity}. The video version can be accessed via the links in the Supplementary Material section or \href{https://www.youtube.com/watch?v=h3b7F8Y8S-g}{this link}.}
\label{fig:3D-K411-lat3-v1} \end{figure}

\begin{figure}
\centering
\includegraphics[width=.99\columnwidth]{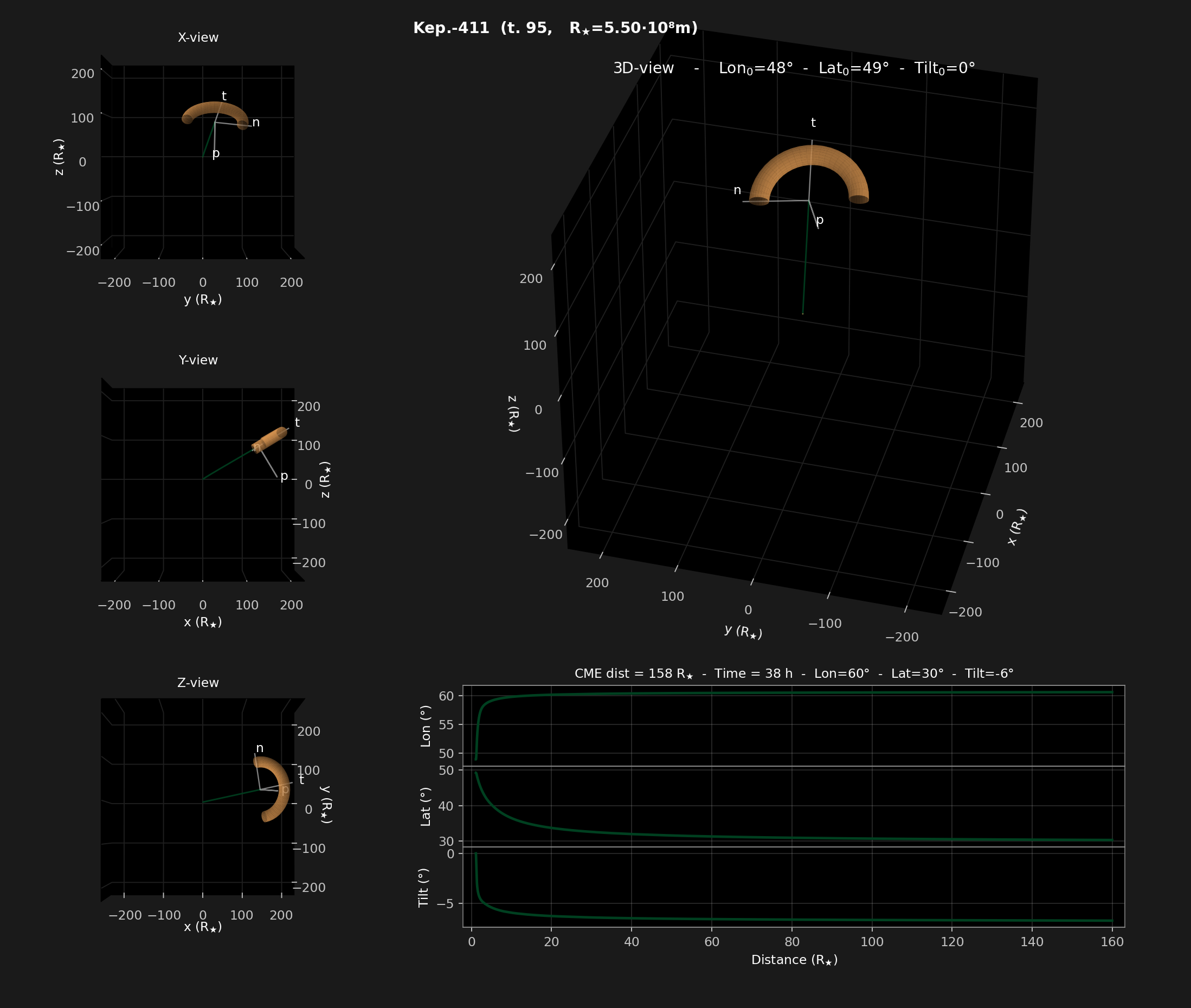}
\caption{Same as Figure~\ref{fig:3D-SUN-lat1-v1}, but for \textbf{Kepler-411} model, \textbf{high-latitude}, \textbf{mid-velocity}. The video version can be accessed via the links in the Supplementary Material section or \href{https://www.youtube.com/watch?v=D4NiSNKgC7w}{this link}.}
\label{fig:3D-K411-lat3-v2} \end{figure}

\begin{figure}
\centering
\includegraphics[width=.99\columnwidth]{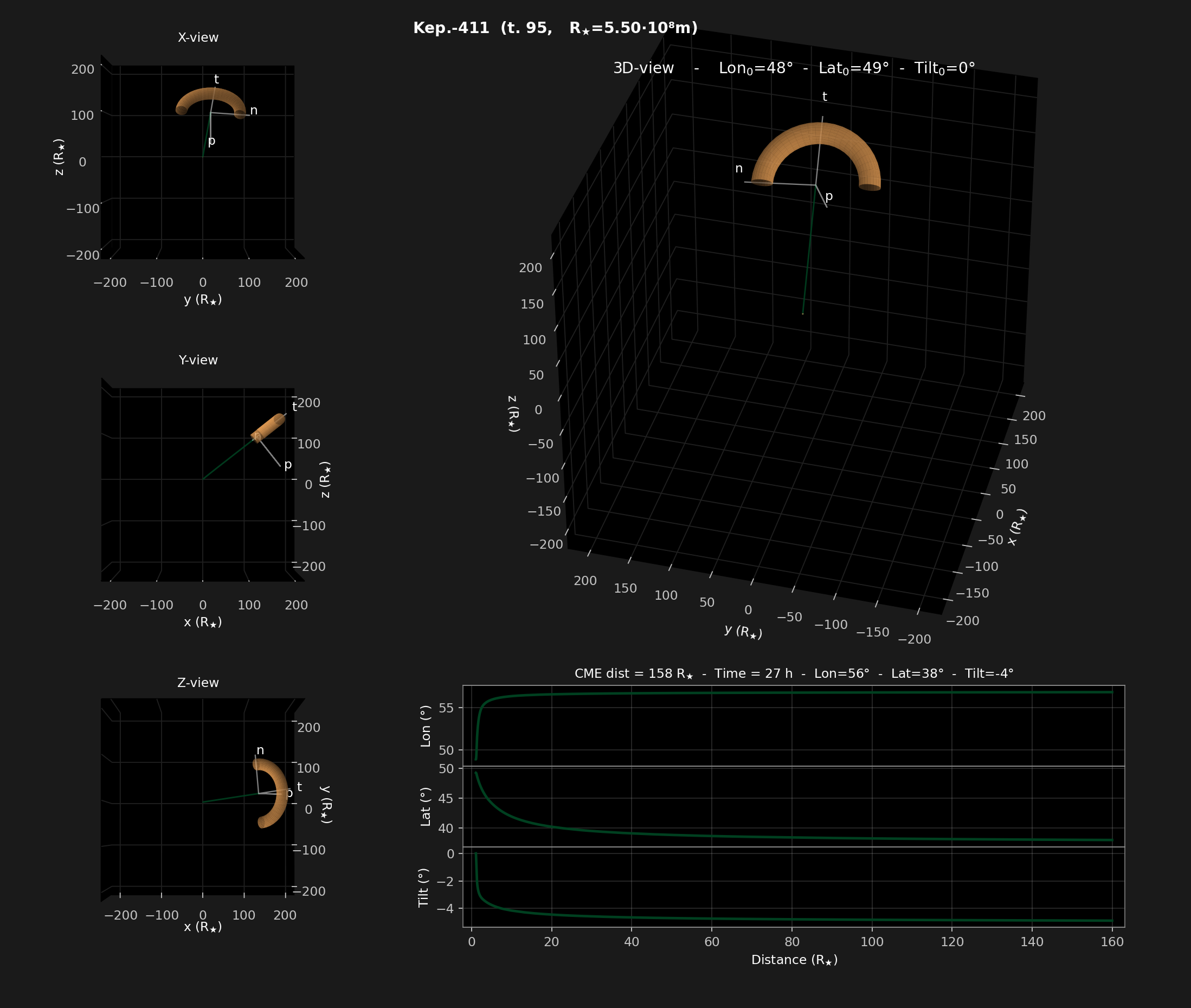}
\caption{Same as Figure~\ref{fig:3D-SUN-lat1-v1}, but for \textbf{Kepler-411} model, \textbf{high-latitude}, \textbf{high-velocity}. The video version can be accessed via the links in the Supplementary Material section or \href{https://www.youtube.com/watch?v=Odtq4c8C70c}{this link}.}
\label{fig:3D-K411-lat3-v3} \end{figure}

\begin{figure}
\centering
\includegraphics[width=\columnwidth]{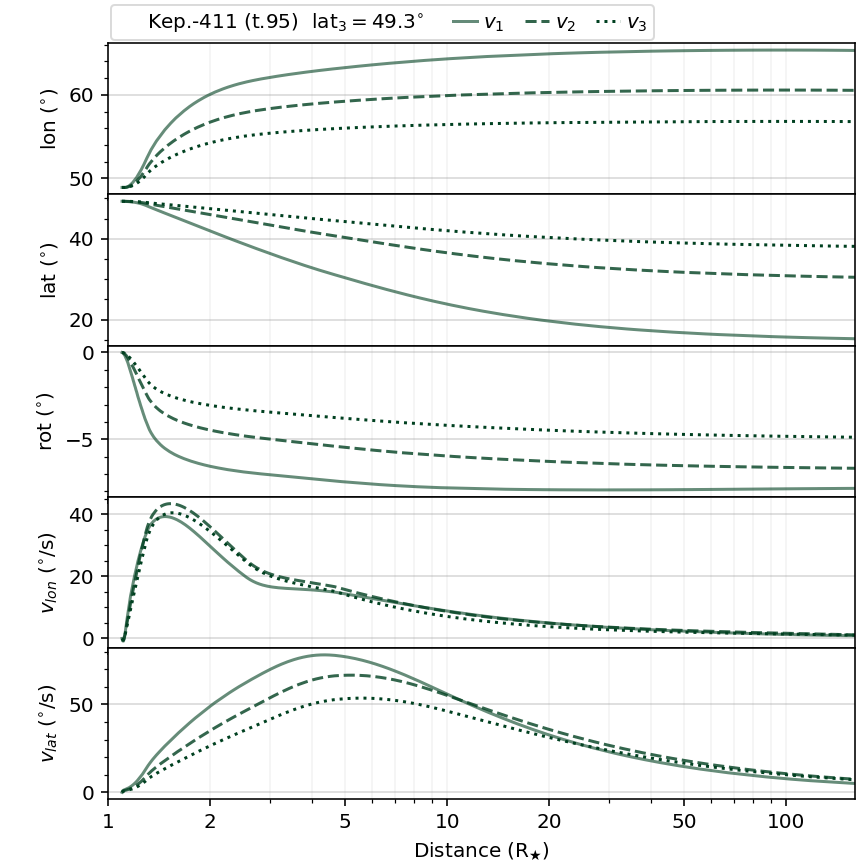}
\caption{Same as Figure~\ref{fig:SUN_l1}, but for \textbf{Kepler-411}, \textbf{high-latitude} simulations.}
\label{fig:K411_l3} \end{figure}

\end{appendices}

\end{document}